\documentclass[fleqn,usenatbib,useAMS]{mnras}

\usepackage{newtxtext,newtxmath}

\usepackage[T1]{fontenc}

\DeclareRobustCommand{\VAN}[3]{#2}
\let\VANthebibliography\thebibliography
\def\thebibliography{\DeclareRobustCommand{\VAN}[3]{##3}\VANthebibliography}

\usepackage{hyperref}
\usepackage{subcaption}
\usepackage{graphicx}	
\usepackage{amsmath}	
\usepackage{tablefootnote}
\usepackage{longtable}
\usepackage{lscape}
\usepackage{orcidlink}

\def\kms{km~s$^{-1}$}
\def\sistff{\ion{Si}{ii}~$\lambda$6\,355\,}

\def\csfe{\ion{C}{ii}~$\lambda$6\,580\,}

\def\phn{\phantom{0}}

\def\farcm{$^{\prime}$}
\def\farcs{$^{\prime\prime}$}

\title[Early light curves of SNe Iax]{SN~2024bfu, SN~2025qe, and the early light curves of type Iax supernovae}

\author[M. R. Magee et al.]{M. R. Magee$^{1}$\orcidlink{0000-0002-0629-8931}\thanks{E-mail: mrmagee.astro@gmail.com}, 
T. L. Killestein$^{1}$, 
M. Pursiainen$^{1}$,
B. Godson$^{1}$,
D. Jarvis$^{2}$\orcidlink{0009-0004-3067-2227},
C. Jim\'enez-Palau$^{3,4}$,
\newauthor
J. D. Lyman$^{1}$,
D. Steeghs$^{1}$,
B. Warwick$^{1}$, 
J. P. Anderson$^{5,6}$\orcidlink{0000-0003-0227-3451},
T. Butterley$^{7}$,
T.-W. Chen$^{8}$, 
V. S. Dhillon$^{2,9}$, 
\newauthor
L. Galbany$^{4,3}$\orcidlink{0000-0002-1296-6887}, 
S. Gonz\'alez-Gait\'an$^{10,5}$\orcidlink{0000-0001-9541-0317},
M. Gromadzki$^{11}$\orcidlink{0000-0002-1650-1518},
C. Inserra$^{12}$, 
L. Kelsey$^{13}$,
A. Kumar$^{14}$\orcidlink{0000-0002-4870-9436},
\newauthor
G. Leloudas$^{15}$\orcidlink{0000-0002-8597-0756},
S. Mattila$^{16,17}$,
S. Moran$^{18}$,
T. E. M\"{u}ller-Bravo$^{19,20}$\orcidlink{0000-0003-3939-7167},
K. Noysena$^{21}$,
G. Ramsay$^{22}$,
\newauthor
S. Srivastav$^{23}$\orcidlink{0000-0003-4524-6883},
R. Starling$^{18}$, 
R. W. Wilson$^{7}$,
D. R. Young$^{24}$\orcidlink{0000–0002–1229–2499},
K. Ackley$^{1}$,
R. P. Breton$^{25}$, 
\newauthor
J. Casares Vel\'azquez$^{9}$, 
M. J. Dyer$^{2}$,
D. K. Galloway$^{26}$, 
E. Kankare$^{16}$, 
R. Kotak$^{16}$,
L. K. Nuttall$^{27}$, 
\newauthor
D. O'Neill$^{28}$,
P. Pessi$^{29}$,
D. Pollacco$^{1}$,
K. Ulaczyk$^{1}$,
O. Yaron$^{30}$ 
\\
\\
\textit{Author affiliations are listed at the end of the paper}
}
\date{Accepted 2025 September 28. Received 2025 September 15; in original form 2025 June 02}

\pubyear{2025}

\begin{document}
\label{firstpage}
\pagerange{\pageref{firstpage}--\pageref{lastpage}}
\maketitle
\begin{abstract}
Type Iax supernovae (SNe Iax) are one of the most common subclasses of thermonuclear supernova and yet their sample size, particularly those observed shortly after explosion, remains relatively small. In this paper we present photometric and spectroscopic observations of two SNe~Iax discovered shortly after explosion, SN~2024bfu and SN~2025qe. Both SNe were observed by multiple all-sky surveys, enabling tight constraints on the moment of first light and the shape of the early light curve. Our observations of SN~2025qe begin \textless2\,d after the estimated time of first light and represent some of the earliest observations of any SN~Iax. Spectra show features consistent with carbon absorption throughout the evolution of SN~2025qe, potentially indicating the presence of unburned material throughout the ejecta. We gather a sample of SNe~Iax observed by ATLAS, GOTO, and ZTF shortly after explosion and measure their rise times and early light curve power-law rise indices. We compare our results to a sample of normal SNe~Ia and find indications that SNe~Iax show systematically shorter rise times, consistent with previous work. We also find some indication that SNe~Iax show systematically lower rise indices than normal SNe~Ia. The low rise indices observed among SNe~Iax are qualitatively consistent with extended $^{56}$Ni distributions and more thoroughly-mixed ejecta compared to normal SNe~Ia, similar to predictions from pure deflagration explosions.
\end{abstract}




\section{Introduction}
\label{sect:intro}

Type Ia supernovae (SNe~Ia) are widely accepted as arising from thermonuclear explosions of white dwarfs in binary systems (see \citealt{liu--23}, \citealt{ruiter--25}, and references therein). The increased cadence and depth of all-sky surveys within the past two decades however has revealed surprising diversity among thermonuclear explosions \citep{dimitriadis--25}. An ever-growing zoo of strange and unusual classes of thermonuclear SNe has been discovered \citep{taubenberger--17}. Based on a volume-limited sample, \cite{dimitriadis--25} estimate that 4.5$\pm$2.5 per cent of SNe~Ia fall into the peculiar type Iax (SN~Iax; \citealt{foley--13, jha--17}) subclass.

\par

Relative to normal SNe~Ia\footnote{We use the term `normal' to indicate `Branch-normal' SNe~Ia and do not consider 91bg- or 91T-like SNe Ia \citep{branch--normal}}, SNe~Iax can be up to seven magnitudes fainter at peak and generally evolve over faster timescales \citep{foley--09, comp--obs--12z, karambelkar--21}. Their near-infrared (NIR) light curves lack the distinct secondary maximum commonly seen in normal SNe~Ia \citep{02cx--orig}, which results from the recombination of iron as the ejecta cool \citep{kasen--06a}. Their spectra are dominated by features due to iron group elements (IGEs) at all epochs \citep{02cx--orig, read--02cx--spectra, 05hk--400days, foley--late--iax}. \cite{foley--13} argue that most, if not all, SNe~Iax show evidence of carbon absorption in their spectra up to maximum light, whereas carbon is observed in only $\sim$30 per cent of normal SNe~Ia \citep{parrent--11, folatelli--12}. Around maximum light, spectra of SNe~Iax typically show lower velocities ($\sim$3\,000 -- 8\,000~\kms; \citealt{foley--13, srivastav--20}) than in normal SNe~Ia. The low luminosity and velocities observed among SNe~Iax are indicative of an overall less energetic explosion compared to normal SNe~Ia, while the apparently ubiquitous presence of carbon indicates incomplete thermonuclear burning is common among the class.

\par

The extreme features of SNe~Iax have resulted in a handful of proposed explosion mechanisms. One scenario that has received considerable attention is the pure deflagration of a Chandrasekhar mass carbon-oxygen white dwarf \citep{read--02cx--spectra, 02cx--late--spec, phillips--07, kromer-13, long--14}. Within this scenario, carbon burning is ignited close to the centre of the white dwarf and propagates as a subsonic, turbulent flame that imparts a significant degree of mixing on the SN ejecta \citep{reinecke--02a, reinecke--02b, fink-2014, lach--22--def}. Overall, predictions from pure deflagration models agree with many of the features that define SNe~Iax \citep{kromer-13, 15h, lach--22--def}, including their low $^{56}$Ni and ejecta masses \citep{fink-2014, mccully--14, lach--22--def, srivastav--22} and the low levels of polarisation observed \citep{chornock--06, bulla--20, maguire--23}. The high level of mixing produced by pure deflagrations is consistent with the level of mixing inferred from spectra of SNe~Iax and the lack of a NIR secondary maximum \citep{phillips--07, 05hk--400days, magee--22}. Nevertheless, notable differences remain -- pure deflagration models typically show stronger carbon features than observed in SNe~Iax \citep{12bwh, magee--22, barna--18, barna--21} and evolve more quickly \citep{kromer-13, 15h,lach--22--def}. A key prediction from the pure deflagration scenario however is that the explosion itself is sufficiently weak that it does not completely unbind the white dwarf, leaving behind a remnant (with masses ranging from $\sim$0.1 -- 1.3~$M_{\odot}$) that may contribute significantly to the SN luminosity, slowing the light curve evolution and improving agreement with observations \citep{kicked--remnants, kromer-13, shen--17, callan--24}.

\par

Rather than a pure deflagration, models invoking a supersonic detonation following an initial deflagration phase have also been proposed \citep{comp--obs--12z}. The transition from deflagration to detonation may be driven by multiple mechanisms, including the collision of deflagration ash around the gravitationally bound white dwarf core or through a series of `pulses' as the core begins to contract following the deflagration \citep{khokhlov--91a, hoeflich-02, plewa--04, lach--22--pdd}. Such detonation models have been argued to be consistent with the velocity evolution and late-time features of SN~2012Z, one of the brightest SNe~Iax observed \citep{comp--obs--12z}, but struggle to reproduce the overall light curves and spectra \citep{lach--22--pdd}. Due to the detonation, the ejecta predicted by these models show layered structures rather than the significant mixing predicted by pure deflagrations. Layering has been argued for at least some SNe~Iax \citep{comp--obs--12z, barna--18, barna--21}, but similarly it has been argued that this is not required to match observations of SNe~Iax \citep{magee--22}.

\par

Both pure deflagrations and deflagration-to-detonation transitions of Chandrasekhar mass carbon-oxygen white dwarfs appear to struggle with reproducing the full diversity of SNe~Iax, particularly the fainter members of the class \citep{fink-2014, lach--22--def, lach--22--pdd}. Alternative scenarios have therefore also been proposed, including explosions of hybrid white dwarfs \citep{meng--14}. \cite{kromer-15} invoke pure deflagrations of hybrid white dwarfs containing a carbon-oxygen core surrounded by an oxygen-neon mantle. This scenario also produces a significant amount of mixing within the ejecta and \cite{kromer-15} argue that synthetic observables are in good agreement with extremely faint SNe~Iax, such as SN~2008ha \citep{foley--09, valenti--09}. \cite{kashyap--18} present a simulation of the merger between an oxygen-neon primary white dwarf and a carbon-oxygen secondary white dwarf. In this scenario, the lower mass secondary becomes disrupted and triggers a carbon detonation resulting in a small amount of $^{56}$Ni and ejecta being released, while a significant amount of the thermonuclear burning products fall back onto the primary. The presence of burning products in the ejecta and fallback material indicates at least some amount of mixing is also found in this scenario. \cite{karambelkar--21} argue in favour of this scenario for SN~2021fcg, one of the faintest SNe~Iax observed to date. 

\par

The distribution of $^{56}$Ni within the ejecta can also provide important insights into the level of mixing among thermonuclear SNe and therefore the explosion mechanism. Following from theory, the early light curves of SNe~Ia have generally been assumed to increase in flux proportional to $t^2$ \citep{riess--99a}. Observational studies have found mean rise indices of $\sim$2 among samples of SNe~Ia, but also significant variation and outliers \citep{gonzalez-gaitan--12, firth--sneia--rise, olling--15, papadogiannakis--2019, miller--20a, fausnaugh--23}. These variations in early light curve shapes have typically been attributed to differences in the $^{56}$Ni distribution in SNe~Ia, with distributions in which at least some $^{56}$Ni extends towards the outer ejecta resulting in brighter light curves at early times and shallower rises towards maximum light \citep{piro-nakar-2013, piro-nakar-2014, noebauer--17, magee--18, magee--20}. Detailed studies of individual SNe~Iax have also revealed shallow, or indeed nearly linear, rises towards maximum light \citep{15h, miller--18, barna--21, maguire--23, hoogendam--25}. Such shallow rises may be further indications of heavy mixing in SNe~Iax, but to date there has not yet been a study to explore the full variation among the class.

\par

In this paper, we present optical photometry and spectroscopy of two recently discovered SNe~Iax: SN~2024bfu and SN~2025qe. Both SNe were observed by multiple surveys and subject to extensive follow up campaigns. Inspired by the constraints on the early light curves enabled through the combination of these surveys, we also investigate the early light curves of a sample of SNe~Iax for the first time. In Sect.~\ref{sect:obs} we discuss the discovery of both SNe and present their observations. Section~\ref{sect:analysis} presents analysis of the light curves and spectra of both SNe, while in Sect.~\ref{sect:early_lcs} we analyse the early light curves of SNe~Iax and SNe~Ia. In Sect.~\ref{sect:discussion} we discuss our results and we present our conclusions in Sect.~\ref{sect:conclusions}.

%

\section{Observations and data reduction}
\label{sect:obs}

\begin{figure*}
\centering
\begin{subfigure}{\columnwidth}
  \centering
  \includegraphics[width=.95\columnwidth]{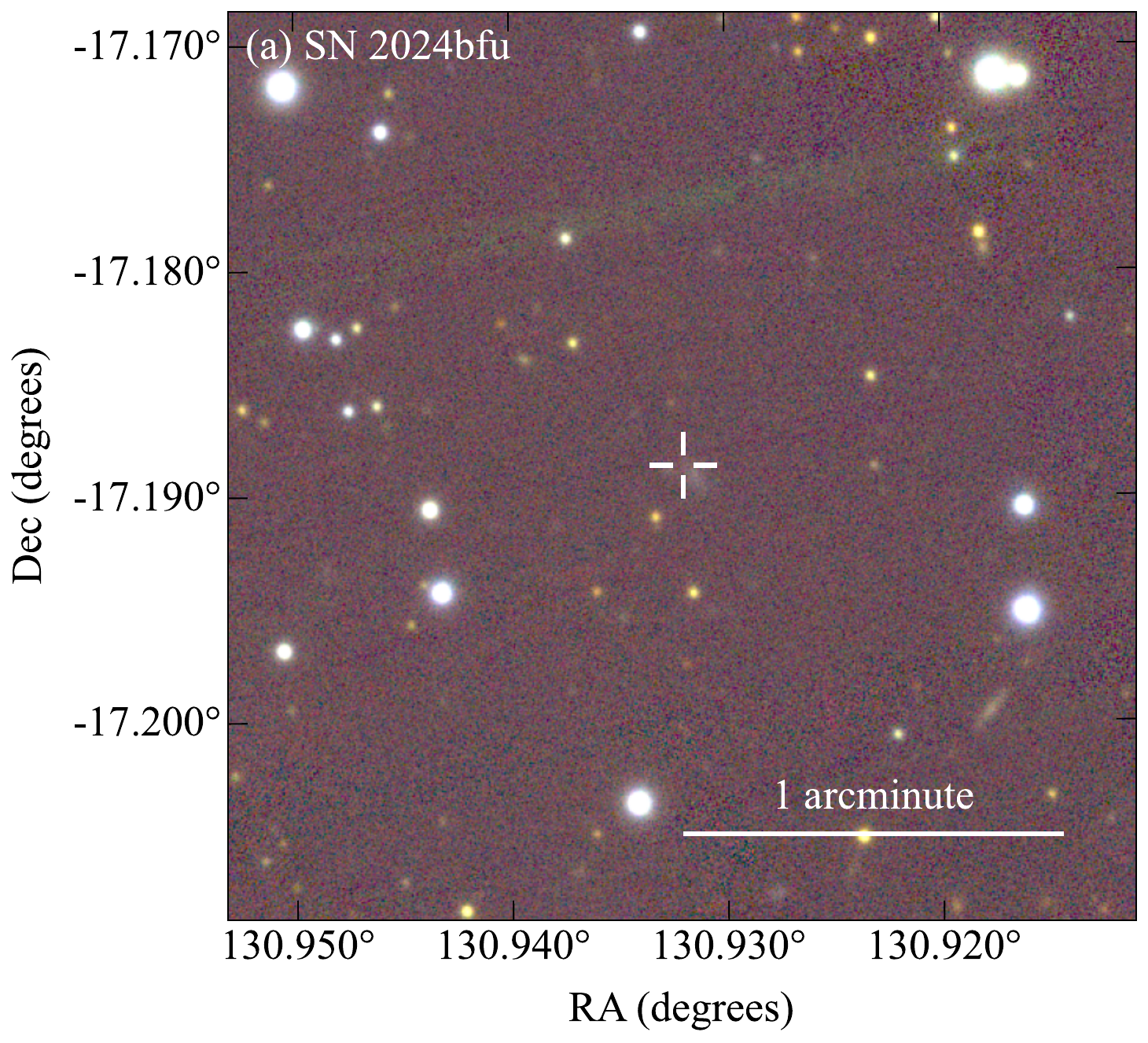}
  \label{fig:sub1}
\end{subfigure}%
\begin{subfigure}{\columnwidth}
  \centering
  \includegraphics[width=.95\columnwidth]{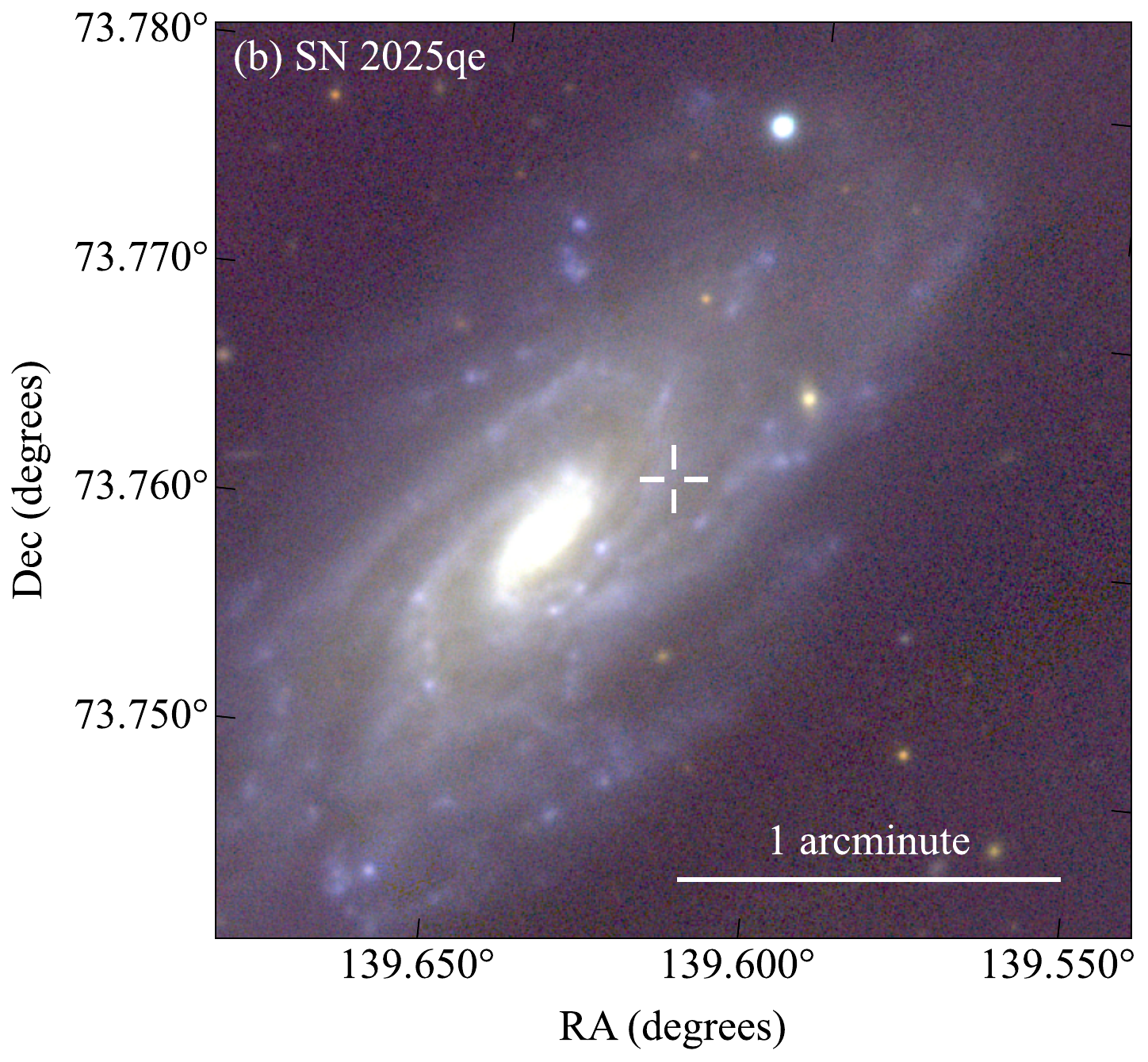}
  \label{fig:sub2}
\end{subfigure}
\caption{Stacked $giz$-band Pan-STARRS images of SN~2024bfu (\textit{Panel a}; $z = 0.036$) and SN~2025qe (\textit{Panel b}; $z = 0.007$). The locations of both SNe are marked by crosshairs.}
\label{fig:finders}
\end{figure*}

SN~2024bfu (Fig.~\ref{fig:finders}(a)) was discovered by the Gravitational-wave Optical Transient Observer (GOTO; \citealt{steeghs--22, dyer--24}) on 2024 Jan. 31 (MJD = 60\,340.56) as part of the GOTO-Fast survey (\citealt{sn2024bfu--discovery}; Godson et al., in prep.) with an apparent magnitude in the GOTO $L$-band (see Sect.~\ref{appendix_filters} in the appendix for discussion of the GOTO $L$-band and filters used in this work) of $m_{L}$ = 18.35$\pm$0.06 \citep{sn2024bfu--discovery}. Subsequent GOTO forced photometry (Jarvis et al., in prep.) revealed that the SN was also observed before discovery with the last non-detection occurring on 2024 Jan. 24 (MJD = 60\,333.54) with a 5$\sigma$ limit of $m_L$ $\textgreater$19.75. In addition, SN~2024bfu was serendipitously observed by the Asteroid Terrestrial-impact Last Alert System (ATLAS; \citealt{atlas, atlas--smith}) prior to discovery, enabling tighter constraints on the explosion epoch. Less than one day after discovery (MJD = 60\,341.30), SN~2024bfu was classified by the extended Public ESO Spectroscopic Survey of Transient Objects+ (ePESSTO+; \citealt{pessto}) as a SN~Iax at a redshift of $z = 0.04$ \citep{sn2024bfu--class}. 

\par

SN~2025qe (Fig.~\ref{fig:finders}(b)) was discovered in IC~0529 by the Zwicky Transient Facility (ZTF; \citealt{bellm--19}) on 2025 Jan. 18 (MJD = 60693.26) with an apparent magnitude in the ZTF $g$-band of $m_{g}$ = 18.67$\pm$0.09 \citep{sn2025qe--discovery}. SN~2025qe was also observed by GOTO approximately six hours earlier with an apparent magnitude of $m_{L}$ = 18.66$\pm$0.19. Following its detection by GOTO, follow up spectroscopic observations were automatically triggered by the GOTO marshall (Lyman et al., in prep.). Across ATLAS, GOTO, and ZTF, the last non-detection of SN~2025qe occurred $\sim$1.7~days before the first detection with a 5$\sigma$ limit of $m_o$ $\textgreater$19.70. Within $\sim$10 hours of the first detection by ZTF, SN~2025qe was observed and classified as a SN~Ia at a redshift of $z = 0.007$ by the Li-Jiang One hour per Night observation of Supernovae survey (LiONS; \citealt{sn2025qe--class, wang--19}), but was later reclassified as a SN~Iax by GOTO \citep{sn2025qe--class--goto}. 

\par

Appendix~\ref{appendix_phot} presents the photometry of SN~2024bfu and SN~2025qe. Photometry from ATLAS, GOTO, and ZTF was obtained using the respective forced photometry services (\citealt{atlas--forced}; Jarvis et al., in prep.; \citealt{masci--19}). For both SNe we remove GOTO epochs for which the SN is within 300 pixels of the CCD edge as these can show additional scatter. Following the method outlined by \cite{rest--25}, we increase the reported flux uncertainties, by estimating scatter in the baseline flux before the SN, to account for additional noise in the data. Supplemental photometry was obtained with IO:O on the Liverpool Telescope (LT; \citealt{lt}) in the $ugriz$ bands, the pt5m \citep{hardy--15} in the $BVRI$ bands, and the 0.8~m telescopes of the Two-meter Twin Telescope (TTT) facility in the $gri$ bands. IO:O images are automatically reduced using the standard LT reduction pipeline. Images from pt5m are reduced for bias, dark, and flat-field corrections using a custom pipeline. TTT data were taken by a CMOS camera and arrived pre-reduced. These data were subsequently median binned over $3\times3$ pixels to remove hot pixels and provide a better sampling of the image quality (unbinned pixels are 0.2\farcs~pix$^{-1}$). Photometry of LT and TTT images was performed using a custom seeing-matched aperture pipeline with difference imaging, and calibrated using a sequence based on Gaia~DR3 synthetic photometry derived from XP spectra~\citep{gaia--synphot}. Template subtraction was performed with \texttt{hotpants}~\citep{becker--hotpants}, using deep reference images from the Panoramic Survey Telescope and Rapid Response System Data Release 1 (Pan-STARRS DR1; \citealt{ps1--chambers}), excluding $u$-band, which was left unsubtracted due to no appropriate templates being available. Photometry of reduced pt5m images was obtained using the photometry-sans-frustration (psf) pipeline \citep{p_s_f}. Template subtraction was also not applied to pt5m images due to the lack of suitable templates.

\par

Appendix~\ref{appendix_spec} gives logs of our spectroscopic observations and instrumental configurations for SN~2024bfu and SN~2025qe. For SN~2024bfu, in total we obtained six epochs of spectroscopy, including two epochs as part of the GOTO-Fast survey and four epochs as part of ePESSTO+. GOTO-Fast spectra were observed using IDS on the Isaac Newton Telescope (INT) and reduced using the PypeIt spectral reduction package \citep{Prochaska2020a, Prochaska2020b} with a custom recipe for IDS. Spectra obtained as part of ePESSTO+ were observed using EFOSC2 on the New Technology Telescope (NTT) and reduced using the PESSTO pipeline \citep{pessto}. For SN~2025qe, we obtained 13 epochs of spectroscopy with SPRAT on the LT and 1 epoch with CAFOS on the CAHA 2.2~m telescope. SPRAT and CAHA spectra were also reduced using PypeIt with custom recipes. Within our analysis we also include the public LiONS classification spectrum of SN~2025qe obtained from the Transient Name Server (TNS)\footnote{\url{https://www.wis-tns.org/}}. All spectra of both SNe were calibrated to an absolute flux level using the observed light curves interpolated to the time of observation.

\begin{table*}
\begin{center}
\caption{Light curve parameters for SN~2024bfu and SN~2025qe.}
\label{tab:peak_fit_results}
\begin{tabular}{ccccccc}
\hline
\textbf{Filter} & \textbf{Rise time} & \textbf{Rise index} & \textbf{Maximum light} & \textbf{Peak apparent} & \textbf{Peak absolute} & \textbf{$\Delta m_{15}$} \\
 & (days) & $\alpha$ &(MJD) & \textbf{magnitude} & \textbf{magnitude} &  (mag) \\

\hline
\hline \multicolumn{7}{c}{\textbf{SN~2024bfu}} \\ 
\hline
$g$ & $\textless$18.41          & $\cdots$                  & $\textless$60\,351.94   & $\textless$18.38 & $\textless$$-$17.85 & $\cdots$ \\[5pt]
$L$ & 14.39$^{+1.68}_{-0.69}$   & 0.86$^{+0.65}_{-0.41}$    & 60\,347.77$\pm$0.25     & 18.15$\pm$0.03 & $-$18.05$\pm$0.15 & $\cdots$ \\[5pt]
$r$ & $\textless$18.41          & $\cdots$                  & $\textless$60\,351.94   & $\textless$17.94 & $\textless$$-$18.23 & $\cdots$ \\[5pt]
$o$ & 19.10$^{+1.71}_{-0.75}$   & 1.24$^{+0.52}_{-0.53}$    & 60\,352.66$\pm$0.38     & 17.85$\pm$0.02 & $-$18.35$\pm$0.15 & 0.70$\pm$0.08 \\[5pt]
$i$ & 23.25$^{+2.58}_{-2.07}$   & $\cdots$                  & 60\,356.96$\pm$1.97     & 17.95$\pm$0.07 & $-$18.18$\pm$0.16 & $\cdots$ \\[5pt]

\hline \multicolumn{7}{c}{\textbf{SN~2025qe}} \\ 
\hline
$u$ & $\textless$12.93 & $\cdots$ & $\textless$60\,704.98 & $\textless$17.30 & $\textless$$-$15.44 & $\cdots$ \\[5pt]
$g$ & 13.35$^{+0.48}_{-0.44}$ & 1.25$^{+0.27}_{-0.32}$ & 60\,705.41$\pm$0.04 & 16.42$\pm$0.03 & $-$16.29$\pm$0.13 & 1.80$\pm$0.04 \\[5pt]
$L$ & 14.13$^{+0.50}_{-0.45}$ & 1.10$^{+0.36}_{-0.34}$ & 60\,706.19$\pm$0.11 & 16.34$\pm$0.01 & $-$16.36$\pm$0.13 & 0.98$\pm$0.07 \\[5pt]
$r$ & 18.86$^{+0.49}_{-0.44}$ & 1.22$^{+0.31}_{-0.29}$ & 60\,710.96$\pm$0.06 & 16.21$\pm$0.01 & $-$16.50$\pm$0.13 & 0.84$\pm$0.02 \\[5pt]
$o$ & 18.12$^{+0.49}_{-0.45}$ & 1.16$^{+0.19}_{-0.18}$ & 60\,710.21$\pm$0.11 & 16.34$\pm$0.01 & $-$16.36$\pm$0.13 & 0.68$\pm$0.02 \\[5pt]
$i$ & 19.89$^{+0.50}_{-0.45}$ & $\cdots$ & 60\,711.99$\pm$0.12 & 16.37$\pm$0.01 & $-$16.31$\pm$0.13 & $\cdots$ \\[5pt]
$z$ & 20.28$^{+0.53}_{-0.49}$ & $\cdots$ & 60\,712.39$\pm$0.22 & 16.49$\pm$0.01 & $-$16.18$\pm$0.13 & $\cdots$ \\[5pt]
\hline
\hline
\end{tabular}
\end{center}
\end{table*}

\subsection{Host galaxies}

SN~2024bfu is most likely associated with a faint, irregular galaxy separated by $\sim$2\farcs\,(see Fig.~\ref{fig:finders}(a)). This source is found in the PS1 catalogue with an apparent magnitude of $m_g$ = 20.78$\pm$0.04, but no spectroscopic redshift is available. The NASA Extragalactic Database (NED) does not contain any galaxies with a spectroscopic redshift within 10\farcm\,of SN~2024bfu. Based on template matching of the classification spectrum using SNID \citep{snid}, SN~2024bfu was originally estimated to have occurred at a redshift of $z = 0.04$ \citep{sn2024bfu--class}. The classification spectrum also shows a weak feature consistent with H$\alpha$ from the host galaxy at a redshift of $z = 0.036$, which we adopt in our analysis. Throughout this work we assume a flat Universe with $H_0$ = 70~km~s$^{-1}$~Mpc$^{-1}$ and $\Omega_\mathrm{M}$ = 0.3, giving a distance modulus of $\mu$ = 36.00$\pm$0.15~mag. 

\par

SN~2025qe is offset by $\sim$22\farcs\,from the centre of the spiral galaxy IC~0529, which has a spectroscopic redshift of $z = 0.007$ \citep{sn2025qe--host--z}. Using the Tully-Fisher distances provided by NED we adopt a weighted mean of $\mu$ = 32.64$\pm$0.13~mag. IC~0529 has a PS1 apparent magnitude of $m_g$ = 11.66$\pm$0.01. 

\par

The host galaxies of both SNe appear typical of SNe~Iax as a whole, which generally occur in actively star-forming galaxies \citep{lyman--13}. The absolute magnitudes of the hosts however show significant differences. For the likely host of SN~2024bfu we find an absolute magnitude of $M_g$ = $-$16.01 and for the host of SN~2025qe we find $M_g$ = $-$21.22. The former is towards the faint end of the distribution for SN host galaxies, while the latter is more typical \citep{hakobyan--12}.

\par

Unlike normal SNe~Ia \citep{lira--relation}, SNe~Iax show significant variation in their colour evolution, making it difficult to estimate extinction due to the host galaxy \citep{foley--13}. We note that none of the spectra of SN~2024bfu or SN~2025qe show strong signs of \ion{Na}{i} absorption, indicating the overall level of host extinction is low (e.g. \citealt{na1d-red-3, na1d-red-1, na1d-red-2}). We therefore do not apply any corrections for host galaxy extinction and instead assume only Milky Way extinction of $A_V = 0.212$~mag and $A_V = 0.074$~mag for SN~2024bfu and SN~2025qe, respectively \citep{schlafly}.

%

\section{Analysis}
\label{sect:analysis}

\subsection{Light curve}
\label{sect:analysis_lc}

\begin{figure*}
\centering
\includegraphics[width=\textwidth]{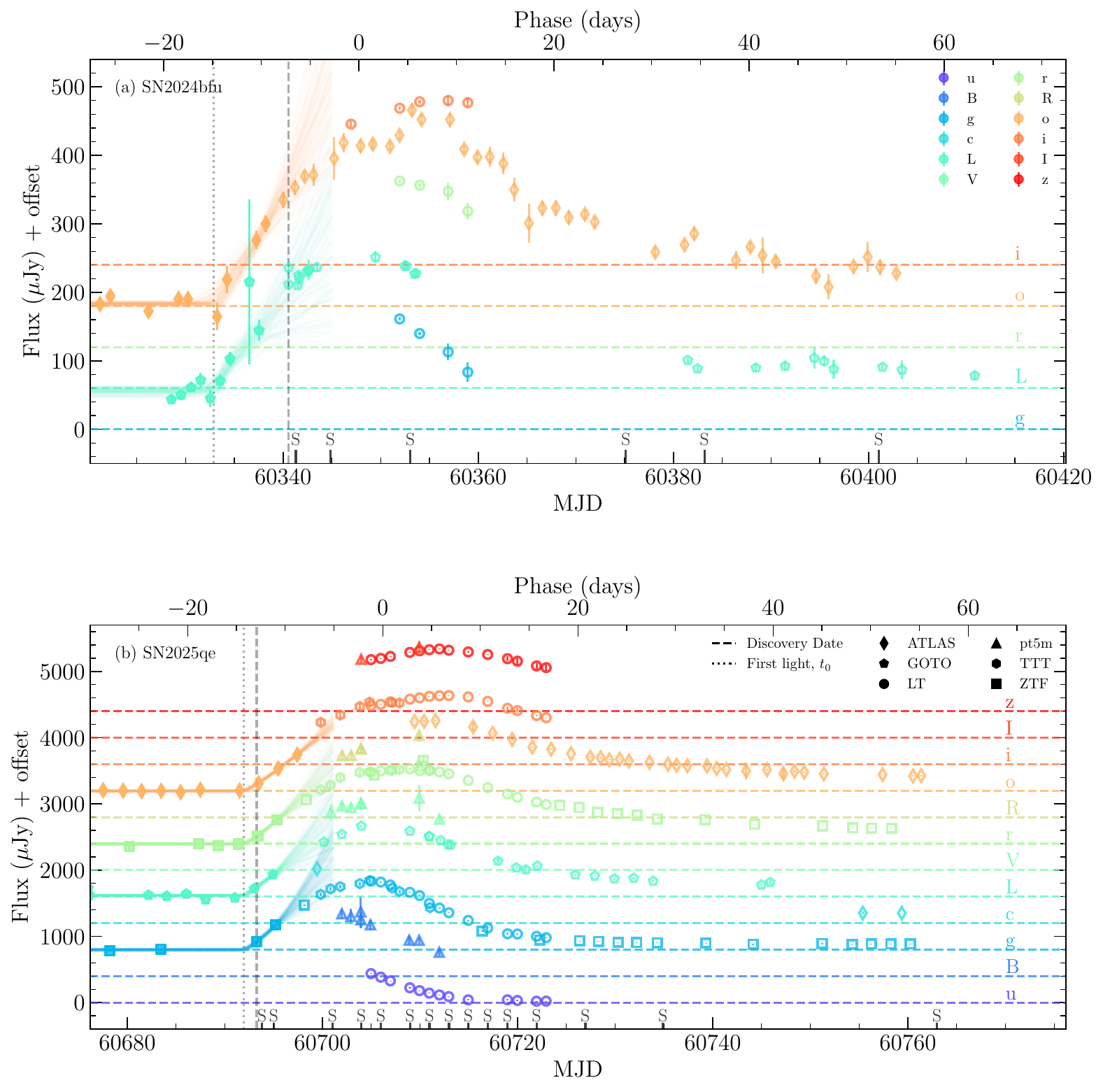}
\caption{Light curves of SN~2024bfu (\textit{Panel a}) and SN~2025qe (\textit{Panel b}). Fits to the observations using Eqn.~\ref{eq:rise} are shown as coloured lines based on randomly sampling the posterior distributions. Observations included in the fits are shown as filled points, while those not included are shown unfilled. Offsets are applied to each filter for clarity and are given by horizontal lines. Observations from different facilities are denoted by different symbols. The epochs of discovery (dashed) and first light (dotted) are marked as vertical lines. Epochs of spectroscopy are denoted by `S'. Phases are given relative to the GOTO $L$-band maximum.}
\label{fig:lc}
\centering
\end{figure*}

\begin{figure*}
\centering
\includegraphics[width=\textwidth]{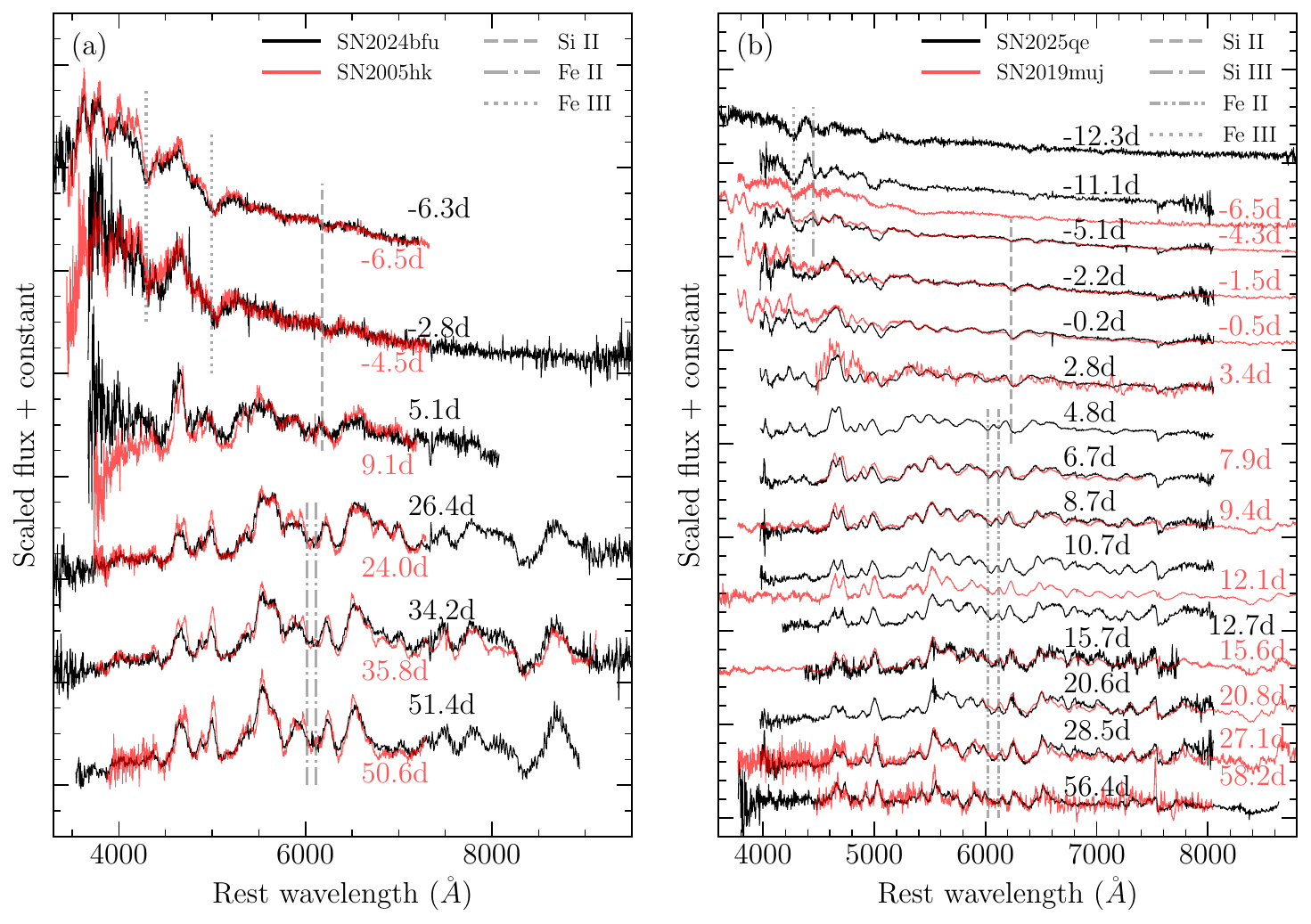}
\caption{Spectra of SN~2024bfu (\textit{Panel a}) and SN~2025qe (\textit{Panel b}) are shown in black. Comparison spectra of SN~2005hk and SN~2019muj are shown in red. Phases of SN~2024bfu and SN~2025qe are given relative to GOTO $L$-band maximum, while SN~2005hk and SN~2019muj are given relative to $V$-band. Spectra have been corrected for Milky Way and host galaxy extinction, where appropriate, and are vertically offset for clarity. Features discussed in the text are marked by vertical lines.}
\label{fig:spec}
\centering
\end{figure*}

Figure~\ref{fig:lc} shows the light curves of SN~2024bfu and SN~2025qe. We perform quadratic fits to each band independently to determine the time of maximum light, peak magnitude, and decline rate in that band. The resulting light curve parameters are given in Table~\ref{tab:peak_fit_results} for both SNe. We find that SN~2024bfu reached a peak absolute magnitude of $M_L$ = $-$18.09$\pm$0.15 on MJD = 60\,347.48$\pm$0.17, which is similar to the well-observed SN~2005hk ($M_V$ = $-$18.07$\pm$0.25; \citealt{comp--obs--12z}) and places it towards the brighter end of the luminosity distribution for SNe~Iax. The ATLAS light curve of SN~2024bfu shows some structure shortly after maximum light (MJD $\sim$ 60\,350 -- 60\,360), with scatter larger than the typical uncertainty despite increased uncertainties to account for additional noise (Sect.~\ref{sect:obs}). We note that all ATLAS observations included here pass the suggested quality cuts, difference images during this time show no significant change, and forced photometry of nearby bright sources shows no increase in scatter in their light curves. It is therefore unclear whether this structure is intrinsic or represents additional, unaccounted for systematic errors. In contrast to SN~2024bfu, SN~2025qe falls into the poorly-sampled region of intermediate-luminosity SNe~Iax with a peak absolute magnitude of $M_L$ = $-$16.36$\pm$0.13, similar to SN~2019muj ($M_V$ = $-$16.42$\pm$0.06; \citealt{barna--21}).

\par

Both SNe were observed serendipitously by multiple surveys throughout their evolutions, enabling us to place tight constraints on the epochs of first light. Following the method outlined by \cite{miller--20a}, we fit the early light curves as a power-law given by:
\begin{equation}
\label{eq:rise}
    f_x(t) = C_x + H[t_{0, x}]A_x(t-t_{0, x})^{\alpha_x},
\end{equation}
where $f_x(t)$ is the flux in band $x$ at time $t$; $C_x$ is a constant representing the baseline, pre-explosion flux in that band; $H[t_{0,x}]$ is the Heaviside function, which is equal to 0 for $t < t_0$ and 1 for $t \geq t_0$; $A_x$ is a proportionality constant in that band; $t_0$ is the time of first light; $\alpha_x$ is the rise index in that band. Following \cite{miller--20a}, we assume a single value of $t_0$ across all bands for a given SN. Although it is unlikely that first light occurs at exactly the same time in all bands, the sampling of our light curves means that in practice we will not be able to resolve any difference between them \citep{magee--18, maguire--23}. In addition, by including only a single value of $t_0$ we are able to simultaneously combine constraints from all bands. We perform the same parameter transformation described by \cite{miller--20a}, whereby $A_x^\prime$ = $A_x$10$^{\alpha_x}$, and fit the light curve up to 50\% of the peak flux, $f \leq 0.5f_{\rm{peak}}$\footnote{As noted by \cite{miller--20a}, the choice of a cut-off threshold is arbitrary and varies within the literature. \cite{miller--20a} apply a threshold of $f \leq 0.4f_{\rm{peak}}$, but show that other values generally produce consistent results within the inferred uncertainties.}. We do not apply any $K$-corrections to our observations given their uncertainty during these early phases and the inhomogeneity of SNe~Iax. Fits are performed using nested sampling with UltraNest \citep{ultranest} and the resulting rise times and indices are given in Table~\ref{tab:peak_fit_results}. Random samples from the posteriors of our UltraNest fits are shown in Fig.~\ref{fig:lc}. 

\par

Based on our fits, we find a median first light time of $t_0$ = 60\,332.87$^{+0.67}_{-1.72}$ for SN~2024bfu. From first light up to maximum, we find rise times of 14.38$^{+1.68}_{-0.69}$~days and 19.10$^{+1.71}_{-0.75}$~days in the GOTO $L$- and ATLAS $o$-bands, respectively. The median power-law indices ($\alpha$) for both bands are $\alpha_L$ = 0.86$^{+0.41}_{-0.65}$ and $\alpha_o$ = 1.24$^{+0.47}_{-0.52}$. While the median rise index of $\alpha_L \textless 1$ is likely unphysical, both indices indicate a shallow rise towards maximum light. These values are generally consistent with rise indices reported for other SNe Iax (see Sect.~\ref{sect:early_lcs}), although we note the uncertainties are relatively large and therefore both the $L$- and $o$-bands are consistent with a single rise index. We note that we also tested the impact of including a shorter baseline before explosion and excluding the last ATLAS $o$-band measurement from our fit. Neither of these changes had a significant impact on our results. After peak, we measure a decline rate in the ATLAS $o$-band of $\Delta m_{15}(o)$ = 0.70$\pm$0.08~mag. We do not attempt to measure a decline rate in the GOTO $L$-band due to data gaps around this phase. 

\par

The luminosity and coverage of SN~2025qe during the early phases enable even tighter constraints of less than one day on the time of first light. We find a first light time of $t_0$ = 60\,691.95$^{+0.44}_{-0.49}$, which is $\textless$11~hours after the previous non-detection by ATLAS on MJD = 60\,691.50. Unfortunately, the peak of the $u$-band light curve was missed and therefore we are unable to determine the $u$-band rise time, but the available data indicates a rise time of $\textless$12.93~days. This is generally comparable to the rise times of other SNe~Iax in the UV, such as SN~2020udy ($t_{u}$ $\lesssim$12\,d; \citealt{maguire--23}) and SN~2019muj ($t_{U}$ = 8.5$\pm$0.7\,d; \citealt{barna--21}). Across the $g$- to $z$-bands, we find longer rise times with increasing wavelength, ranging from approximately two weeks in the $g$-band ($t_{g}$ = 13.35$^{+0.48}_{-0.44}$~days) to three weeks in the $z$-band ($t_{z}$ = 20.28$^{+0.53}_{-0.49}$~days). The median power-law indices across all bands range from $\sim$1.1 -- 1.3, again indicating a relatively shallow rise in all bands and consistent with other SNe~Iax (Sect.~\ref{sect:early_lcs}). As was the case for SN~2024bfu, the relatively large uncertainties on $\alpha$ mean that we are unable to determine whether SN~2025qe shows any trend in $\alpha$ across wavelength -- all bands are consistent with a single, low $\alpha$. After peak, we find slower decline rates with increasing wavelength, ranging from $\Delta m_{15}(g)$ = 1.80$\pm$0.04~mag to $\Delta m_{15}(o)$ = 0.68$\pm$0.02~mag. Again, this is comparable to SN~2019muj, which showed decline rates of $\sim$2.0 -- 0.7 in the $g$- to $i$-bands \citep{barna--21}.

\subsection{Spectra}

In Fig.~\ref{fig:spec} we show the spectra of SN~2024bfu and SN~2025qe. In both cases, our spectra cover the optical evolution from pre-maximum to a few weeks after maximum light. In particular, our spectral sequence for SN~2025qe includes some of the earliest observations of any SN~Iax at 1.8\,d and 3.0\,d after the estimated time of first light. SN~2020udy was observed at 2.3\,d after first light, but was $\textgreater$1~mag brighter than SN~2025qe \citep{maguire--23}. The combined photometric and spectroscopic sequence of SN~2025qe therefore offers a unique look at the early-time properties of SNe~Iax and in particular intermediate-luminosity or faint members of the class. Figure~\ref{fig:spec} also includes comparison SNe~Iax with comparable luminosities, SN~2005hk \citep{phillips--07, blondin--12} and SN~2019muj \citep{barna--21, kawabata--21}, at similar phases. All comparison spectra were obtained from WISeREP \citep{wiserep} and corrected for Milky Way (and host) extinction where appropriate. Velocities of spectral features are measured based on Gaussian profile fits to the continuum-normalised feature, with uncertainties obtained from sampling of the posterior distributions. 

\par

Figure~\ref{fig:spec}(a) demonstrates that SN~2024bfu shows remarkable similarities to SN~2005hk throughout its pre- to post-maximum evolution. The earliest spectrum of SN~2024bfu at $-$6.3\,d shows features typical of bright SNe~Iax at this epoch -- a blue continuum dominated by strong \ion{Fe}{iii} absorption and weak intermediate mass elements (IMEs). Such weak features make robust identifications and velocity measurements challenging, but we find a feature at $\sim$6\,200~\AA\, that is consistent with \sistff blueshifted by 8\,300$\pm$1\,400~\kms. This is comparable to the velocities of other SNe~Iax measured at similar phases \citep{foley--13} and the velocities we find for the strong \ion{Fe}{iii}~$\lambda$4\,404 (7\,600$\pm$200~\kms) and $\lambda$5\,129 (7\,700$\pm$400~\kms) absorption features. In our $-$2.8\,d spectrum, the likely \sistff feature has increased in strength and decreased in velocity to 4\,100$\pm$500~\kms. By $+$5.1\,d this feature has further increased in strength and broadened, likely indicating that it is now blended with \ion{Fe}{ii} \citep{magee--22}, but the relatively low signal-to-noise ratio of the spectrum means we are unable to robustly identify multiple absorption troughs. The $+$5.1\,d spectrum also shows that prominent \ion{Fe}{ii}~$\lambda$6\,149 and $\lambda$6\,247 absorption features have appeared, similar to the $+$9.1\,d spectrum of SN~2005hk, further supporting an \ion{Fe}{ii} contribution to the feature at $\sim$6\,300~\AA. Progressing to later times, the spectra of SN~2024bfu follow the typical evolution of SNe~Iax, becoming increasingly dominated by features due to iron group elements (IGEs) and showing a strong \ion{Ca}{ii}~NIR triplet. While SN~2024bfu continues to show strong similarities with SN~2005hk throughout its evolution, beginning with our $+26.4$\,d spectrum one of the most notable differences relative to SN~2005hk occurs in the wavelength region $\sim$5\,700 -- 6\,000~\AA. Within this region, SN~2005hk shows two absorption features at $\sim$5\,750~\AA\, and $\sim$5\,900~\AA, which have been attributed to \ion{Na}{i} and \ion{Co}{ii}, respectively \citep{read--02cx--spectra, 05hk--400days}. SN~2024bfu however does not show any significant absorption features within this region and instead shows a relatively flat pseudo-continuum (particularly in the $+26.4$\,d spectrum). The \ion{Na}{i} feature has begun to develop in our $+$34.2\,d spectrum. By $+51.4$\,d it is now clearly visible with a velocity of 5\,700$\pm$200~\kms\,(comparable to the velocities of \ion{Fe}{ii}~$\lambda$6\,149 and $\lambda$6\,247), but the weaker feature at $\sim$5\,900~\AA\, remains absent. We do not find any strong evidence in favour of \ion{He}{i} absorption features, but we note that we cannot rule out the presence of helium within the ejecta \citep{magee--19}.

\par

\begin{figure}
\centering
\includegraphics[width=\columnwidth]{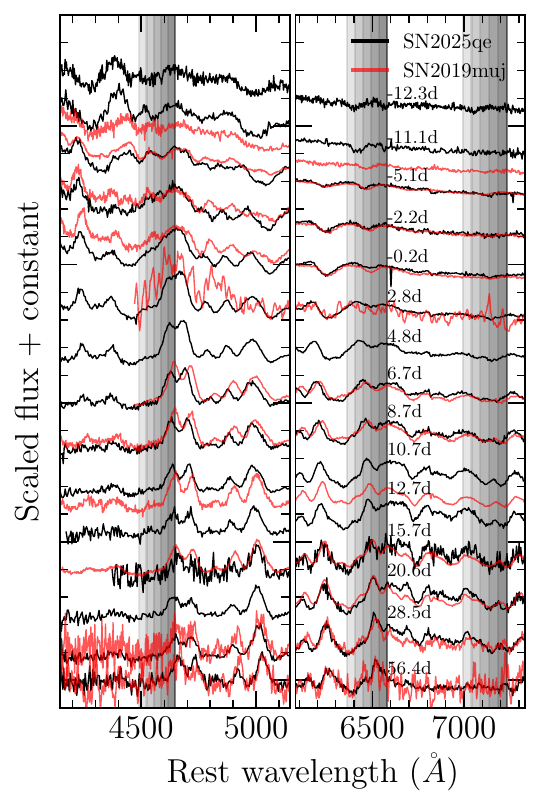}
\caption{Zoom-in of the region surrounding \ion{C}{iii}~$\lambda$4\,647 (\textit{left}) and \ion{C}{ii}~$\lambda$6\,580 \& $\lambda$7234 (\textit{right}) for SN~2025qe (black) and SN~2019muj (red). Shaded regions indicate velocity offsets in steps of 2\,000~\kms.}
\label{fig:spec_cii}
\centering
\end{figure}

As shown in Fig.~\ref{fig:spec}(b), the spectra of SN~2025qe display many similarities to SN~2019muj. The $-$12.3\,d and $-$11.1\,d spectra of SN~2025qe represent some of the earliest spectral observations of any SN~Iax and therefore no spectra of SN~2019muj exist at comparable phases. As with SN~2024bfu, the early spectra of SN~2025qe show the characteristic blue continuum, weak features due to IMEs, and strong \ion{Fe}{iii} absorption of young SNe~Iax. Unlike SN~2024bfu however, both spectra also show strong absorption due to \csfe, indicating unburned material must be present in the outermost layers of the ejecta, with no obvious signs of \sistff at these epochs. Indeed, absorption features consistent with \csfe, and potentially other carbon features, persist throughout the spectral evolution up to maximum light and weeks later, similar to SN~2019muj although with somewhat higher velocities. 

\par

In Fig.~\ref{fig:spec_cii} we show a zoom-in of the wavelength regions surrounding tentative carbon features. \ion{C}{iii}~$\lambda$4\,647 has been potentially identified in a handful of SNe~Iax spectra during the very earliest phases and our earliest spectra show an absorption feature at $\sim$4\,550~\AA\, that is consistent with this identification. From our $-$12.3\,d spectrum we measure a velocity of 7\,400$\pm$600~\kms. This is consistent with the velocities measured for the \ion{Fe}{iii}~$\lambda$4\,404 (8\,700$\pm$700~\kms) and \ion{Si}{iii}~$\lambda$4\,553 (6\,800$\pm$200~\kms) absorption features at $\sim$4\,270~\AA\, and $\sim$4\,450~\AA, respectively. At $-$5.1\,d the \ion{C}{iii} velocity decreases to 4\,000$\pm$1\,000~\kms, again consistent with \ion{Fe}{iii}~$\lambda$4\,404 (5\,000$\pm$600~\kms) and \ion{Si}{iii}~$\lambda$4\,553 (4\,000$\pm$400~\kms). Similar features are also observed in SN~2019muj, although in this case these persist until approximately maximum light and show little evolution in later spectra. At $-$12.3\,d, we measure a velocity of 8\,300$\pm$100~\kms\, for \csfe, which rapidly drops to 7\,400$\pm$100~\kms\, at $-$11.1\,d and 5\,900$\pm$200~\kms\, at $-$5.1\,d. These velocities are also comparable to, albeit slightly higher than, the potential \ion{C}{iii}~$\lambda$4\,647 features identified. By maximum light, the \csfe velocity has decreased by a factor of $\sim$2, compared to the $-$12.3\,d, spectrum to 4\,200$\pm$800~\kms. At this phase, a feature consistent with \ion{C}{ii}~$\lambda$7\,234 begins to appear with a similar velocity. This feature is not clearly apparent in SN~2019muj at a similar epoch. Post-maximum the difference in the \csfe velocity relative to SN~2019muj becomes more apparent, with the former showing a velocity of 3\,300$\pm$300~\kms\, more than one week after maximum light. By our $+28.5$\,d spectrum, the \csfe feature has decreased to a velocity of $\sim$800$\pm$400~\kms, which is lower than the velocities we measure for \ion{Fe}{ii}~$\lambda$6\,149 \& $\lambda$6\,247 at this epoch (3\,700$\pm$200~\kms\, \& 3\,500$\pm$200~\kms, respectively). Low carbon velocities relative to other species have been identified in other SNe~Iax \citep{comp--obs--12z, tomasella--2016, tomasella--20}. At $+56.4$\,d, the \csfe feature is no longer visible. Section~\ref{sect:carbon} provides further discussion of the identification of carbon features and implications for explosion models.

\par

In addition to higher velocity \ion{C}{ii} features, the pre-maximum spectra of SN~2025qe also show generally higher velocity (and stronger) features due to IGEs (in particular \ion{Fe}{iii}~$\lambda$4\,404 and $\lambda$5\,129) and a somewhat redder continuum than observed in SN~2019muj at similar phases (Fig.~\ref{fig:spec}). Indeed, we find the $g-r$ colour of SN~2025qe is $\sim$0.2~mag redder than SN~2019muj up to shortly after maximum light. The \sistff feature has emerged in SN~2025qe by the time of our $-5.1$\,d spectrum and persists up to maximum light, showing a slight velocity decrease from 5\,500$\pm$300~\kms\, to 5\,100$\pm$100~\kms\, at $-$0.2\,d (comparable to the velocities we measure for the \ion{C}{ii} features). This feature continues to broaden post-maximum, again indicating increased contamination from \ion{Fe}{ii} absorption. The $+6.7$\,d and $+8.7$\,d spectra of SN~2025qe show a broad and complex absorption feature centred around $\sim$6\,300~\AA\, that is both higher velocity than in SN~2019muj and has a stronger \ion{Fe}{ii} component. The \ion{Fe}{ii}~$\lambda$6\,149 and $\lambda$6\,247 features are also notably stronger and higher velocity than SN~2019muj at this phase. Again as with SN~2024bfu, SN~2025qe becomes increasingly dominated by features due to IGE at later epochs and we find no evidence in favour of \ion{He}{i} absorption, but cannot rule out the presence of helium in the ejecta.

%

\section{Early light curves}
\label{sect:early_lcs}

SN~2024bfu and SN~2025qe were both observed by multiple surveys around the time of first light. Inspired by this coverage, we select a sample of SNe~Iax with similar or better coverage around these times, perform fits to their early light curves, and analyse the properties of SNe~Iax as a class. We also include a sample of normal SNe~Ia analysed in the same way as reference.

\par

We begin by selecting SNe discovered from 2018 Jan 01 to 2025 Feb 01 and publicly classified on the TNS as `SN~Iax'. In total, this initial sample includes 44 SNe~Iax. As our sample is based on public spectroscopic classifications it is inevitably incomplete. Many SNe~Iax are likely not spectroscopically classified or not classified explicitly as `SN~Iax'. Instead they may be classified as `SN~Ia-pec', as in the case of SN~2024bfu, or indeed even `SN~Ia', as in the case of SN~2025qe initially. Nevertheless, by limiting our analysis to only those objects that have been spectroscopically classified as SN~Iax we reduce the impact of contamination from other classes, including other peculiar SNe~Ia. Light curves of this sample were obtained from the ATLAS, GOTO, and ZTF forced photometry services, including quality cuts and increases to flux uncertainties outlined in Sect.~\ref{sect:obs}. We fit the light curves of each SN following the method described in Sect.~\ref{sect:analysis_lc}. Our final sample includes only those SNe that have at least four detections in a given band included in the fit after the estimated time of first light (i.e. four detections wherein the flux $f \leq 0.5f_{\rm{peak}}$). This selection cut is made to limit any potential bias towards low (or indeed linear) $\alpha$ arising from fitting only two or three data points. Finally, to further reduce contamination we remove any SNe with peak absolute magnitudes $M \textless -18.5$ (SN~2021afcp). In total this gives us a final sample of 14 SNe~Iax. The results of our fits are given in Appendix~\ref{appendix_props}.

\par

To produce a reference sample of SNe~Ia we select all objects classified on the TNS by ePESSTO+ within the past two years as `SN~Ia' up to $z \leq 0.08$ -- totalling 323 SNe. By selecting our sample from only those SNe classified by ePESSTO+ we ensure a uniform sample in which all classification spectra were obtained with the same instrument and classified in the same way, again limiting the possibility of contamination from other peculiar thermonuclear SNe that were, for various reasons, classified and announced simply as `SN~Ia'. We note that by selecting SNe~Ia classified by ePESSTO+, most of our targets occurred in the southern hemisphere. This introduces a bias towards redder filters (such as the ATLAS $o$-band) as only $\sim$25 per cent of our sample were observed by ZTF and therefore most lack the bluest band (ZTF $g$-band). The choice of a redshift limit of $z \leq 0.08$ was also made to ensure our SNe~Ia sample covers a similar redshift range as our SNe~Iax sample. Light curves were again obtained from the survey forced photometry services. To further reduce contamination and remove over- or under-luminous SNe~Ia sub-types, we apply a luminosity cut and select only those SNe~Ia for which the peak magnitude is $-18.5 \leq M \leq -20$. Applying the same criteria of requiring four detections included in the fit after the estimated time of first light, this gives a final sample of 87 SNe~Ia.

\par

We note that for both SNe~Ia and SNe~Iax we find no significant correlation between the number of detections and the rise index. Indeed, SN~2018cxk has some of the lowest rise indices ($\alpha$ $\sim$ 1) in both the $g$- and $r$-bands despite having $\textgreater$20 detections in each band -- more than any other SN in our sample (Fig.~\ref{fig:fit_18cxk}). These values are also in agreement with the rise indices found by \cite{miller--20a}. We also note that \cite{miller--20a} and \cite{fausnaugh--23} report an observational bias in which the rise index and rise time are both correlated with redshift -- higher redshift (fainter) SNe are discovered later and biased towards smaller rise times. These correlations disappear when limiting the sample to the highest quality SNe. Similar to \cite{miller--20a} and \cite{fausnaugh--23}, we find no similar correlation, which is likely related to the selection criteria (public classification spectrum), quality cuts (number of detections), and redshift limit (SNe~Ia with $z \leq 0.08$) used for our sample.

\begin{figure*}
\centering
\includegraphics[width=\textwidth]{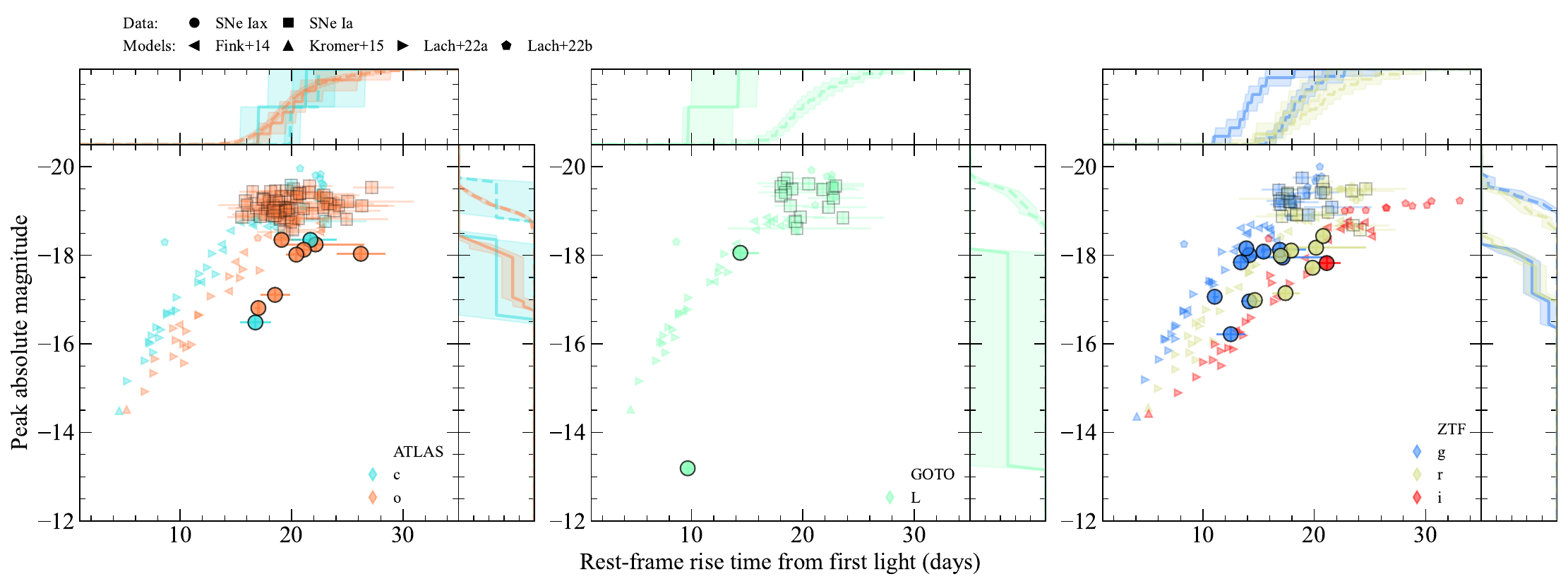}
\caption{Peak absolute magnitudes and rise times for normal SNe~Ia (squares) and SNe~Iax (circles). Predictions from pure deflagration models \citep{fink-2014, kromer-15, lach--22--def} are shown as triangles, while predictions from pulsationally assisted gravitationally confined detonations are shown as pentagons \citep{lach--22--pdd}. Each band is given by a different colour. Cumulative-density histograms for peak absolute magnitude and rise time are shown in the top and right-hand side panels, respectively. Within these panels, normal SNe~Ia are given by a dashed line, while SNe~Iax are given by a solid line. Distributions are shown based on randomly sampling the SNe within each band and the posteriors of our UltraNest fits 10\,000 times, with prominent lines showing the median and shaded regions showing the 1$\sigma$ deviation.}
\label{fig:abs_mag_vs_rise_time}
\centering
\end{figure*}

\begin{figure*}
\centering
\includegraphics[width=\textwidth]{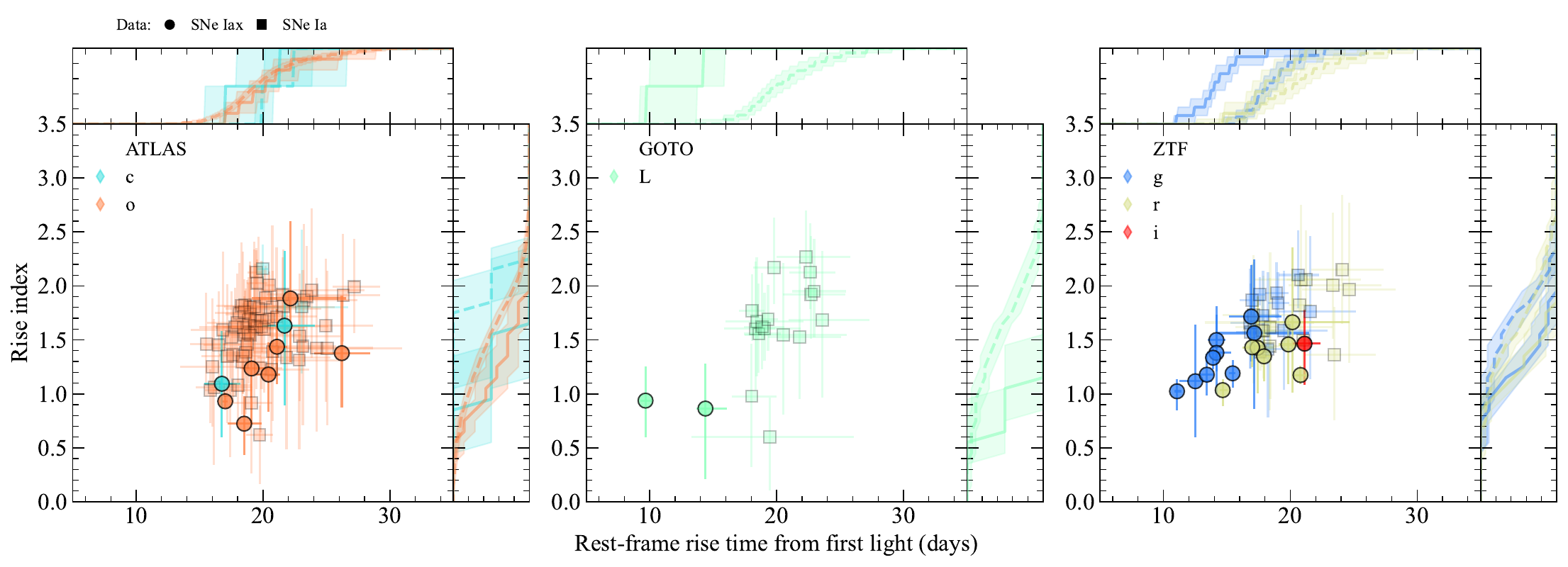}
\caption{As in Fig.~\ref{fig:abs_mag_vs_rise_time} for the rise index and rise time of SNe in our samples.}
\label{fig:rise_index_vs_rise_time}
\centering
\end{figure*}

In Fig.~\ref{fig:abs_mag_vs_rise_time} we present the distributions of peak absolute magnitudes and rise times for our samples of SNe~Iax and normal SNe~Ia. Figure~\ref{fig:abs_mag_vs_rise_time} shows the broad range of peak absolute magnitudes expected of SNe~Iax, which are systematically fainter than normal SNe~Ia. Using a sample of SNe~Ia observed by ZTF as part of a high cadence survey, \cite{miller--20a} found a broad range of rise times, with no evidence in favour of a single rise time, for SNe~Ia. We find that the rise times of SNe~Iax show a similarly broad and heterogeneous distribution. For the extremely faint SN~2024vjm ($M_L$ = $-$13.19$\pm$0.15) we measure a rise time of $t_{L}$ = 9.66$^{+0.47}_{-0.40}$\,d in the GOTO $L$-band, while the much brighter SN~2018cxk ($M_g$ = $-$17.06$\pm$0.15) has a similar short rise time of $t_{g}$ = 11.06$^{+0.34}_{-0.18}$\,d in the ZTF $g$-band. In general, our results show that the rise times of SNe~Iax may be systematically shorter than normal SNe~Ia, which is consistent with previous studies \citep{phillips--07, foley--09, obs--07qd, foley--13}. The difference between rise times of SNe~Iax and normal SNe~Ia is most striking in the bluer bands (ZTF $g$- and GOTO $L$-bands), while we find a trend of longer rise times in redder bands. Robustly assessing the statistical significance of these trends however requires a larger homogeneous sample of both SNe~Iax and normal SNe~Ia.

\par

Based on rise time measurements of a literature sample of SNe~Iax, \cite{15h} suggested the possibility of a correlation between peak absolute magnitude and rise time in the $r$/$R$-bands, with brighter SNe~Iax typically taking longer to reach maximum. We investigate the possibility and significance of such a correlation among our SNe~Iax by randomly sampling the population and posteriors 10\,000 times. From this we find a median Spearman correlation coefficient of $-$0.64 for the ZTF $r$-band, indicating a strong correlation. This is comparable to the correlation coefficients found by \cite{15h} ($\sim$ $-$0.5 -- $-$0.7 depending on which objects are included). Unfortunately however, the smaller sample size (7 SNe~Iax cf. 12 in \citealt{15h}) and relatively large measurement uncertainties mean that this is less statistically significant (median $p$-value 0.12 cf. $\sim$0.02 -- 0.08 in \citealt{15h}). We perform the same test for the ZTF $g$- and ATLAS $o$-bands and find similar correlations ($\sim$$-$0.50 and $-$0.43, respectively), but again with low significance. We note that the ATLAS $c$- and GOTO $L$-band samples contain only two SNe~Iax. Our results therefore generally support the correlation reported by \cite{15h}, but larger samples and more precise rise time measurements are needed. We find no similar correlations between peak magnitude and rise time among SNe~Ia in any band (correlation coefficients $\sim$0), but we note that our SNe~Ia sample does not include fainter sub-types such as 91bg-like SNe. We also note that our SNe~Iax sample is heavily skewed towards the brighter end of the distribution, with only two SNe in our sample having peak magnitudes fainter than $-$16.5 (compared to approximately half in \citealt{15h}). Given that these bright SNe~Iax tend to have longer rise times, this will also bias our rise time distributions and therefore impact whether SNe~Iax are statistically different as a whole compared to normal SNe~Ia. A more representative distribution of SNe~Iax may show larger differences between the two populations.

\par

Figure~\ref{fig:rise_index_vs_rise_time} shows the rise indices measured for our SNe~Iax and normal SNe~Ia samples against their respective rise times. \cite{miller--20a} find a mean rise index of $\sim$2 in both the ZTF $g$- and $r$-bands, but also significant variation with rise indices extending from $\sim$0.5 -- 2.5. Based on our fits, we find similar means and spreads in rise indices for normal SNe~Ia. Likewise, our SNe~Iax sample also shows a broad range of rise indices. We find some evidence in favour of SNe~Iax rise indices being systematically lower than normal SNe~Ia. Indeed, our fits indicate weighted mean rise indices of $\sim$1.4 -- 1.5 in the ZTF $g$- and $r$-bands for SNe~Iax. As with the rise times, the ZTF $g$- and GOTO $L$-bands show the most prominent differences relative to the normal SNe~Ia sample. Again we note however that robustly testing the statistical significance of this requires a larger and more homogeneous sample.

\par

\cite{miller--20a} argue in favour of a correlation between rise time and rise index for their sample of 51 SNe~Ia, with those SNe~Ia having longer rise times also showing sharper rises (higher $\alpha$) towards maximum light. Based on our sample of 11 and 15 SNe~Ia in the ZTF $g$- and $r$-bands, respectively, with both rise time and rise index measurements we find no strong evidence for a similar correlation. This likely arises from both the smaller sample used in this work and the overall poorer sampling of our light curves relative to the high cadence sampled used by \cite{miller--20a} resulting in larger uncertainties, particularly on the rise index. While our rise time measurements benefit from the combined constraints on first light across all bands, the precision of the rise index is still limited by the cadence in each band. \cite{miller--20a} also include some SNe~Ia at the extremes of the luminosity or stretch distribution that would be removed by our magnitude cut, which could also contribute to the differences found between samples. Similarly, we do not find evidence in favour of a correlation in either the GOTO $L$-band or ATLAS bands. For our SNe~Iax sample we find the ZTF $g$-band and ATLAS $o$-band show weak evidence of correlations (coefficients $\sim$0.4), but these are not statistically significant. Similarly, SNe~Iax show some correlations between the peak absolute magnitude and rise index, with brighter SNe typically having larger rise indices, but none are statistically significant. Normal SNe~Ia show no correlation.

\par

In summary, we find a broad range of rise times and rise indices among both SNe~Iax and normal SNe~Ia. Overall, our results highlight the need for increased spectroscopic classification and higher cadence surveys. Across seven years of observations, only 44 SNe~Iax were spectroscopically classified as such and publicly announced on the TNS. It is likely however that at least some SNe~Iax have been misidentified or not recognised. The inclusion of light curve and velocity information during classification, rather than purely template matching, would aid in producing more complete samples and also subsequently increasing the number of templates. Furthermore, even with light curves combined across three different surveys (each one with a typical cadence of $\sim$2 -- 3~days), only 14 of these SNe had sufficient data (i.e. four detections after first light and before 0.5$f_{\rm{peak}}$) to be included in our analysis. The combined light curve quality also varied greatly, limiting our ability to assess statistical trends. Future surveys with higher cadence are required to fully assess the significance of the trends identified in our analysis, particularly in redder bands. Alternatively, a consistent set of filters across multiple surveys could potentially achieve a similarly higher cadence or better sampled light curve. More work and larger samples may show whether the correlations and trends identified in this and previous works are significant or insignificant.

%

\section{Discussion}
\label{sect:discussion}

\subsection{Carbon in SN~2025qe}
\label{sect:carbon}

The $-$12.3\,d and $-$11.1\,d spectra of SN~2025qe represent some of the earliest spectral observations of any SN~Iax, occurring only 1.8\,d and 3.0\,d after the estimated time of first light. Both spectra (and indeed most spectra of SN~2025qe) show features consistent with carbon absorption. Carbon features have been at least tentatively identified in a number of SNe~Iax, including \ion{C}{ii} and \ion{C}{iii} \citep{tomasella--2016, 12bwh, srivastav--20, tomasella--20, barna--21, srivastav--22, maguire--23, hoogendam--25}. Indeed, \cite{foley--13} argue that every SN~Iax with a spectrum before or around maximum shows some indication of carbon absorption.

\par

In our $+$28.5\,d spectrum of SN~2025qe, we identify a feature consistent with \ion{C}{ii}~$\lambda$6\,580 at $\sim$800$\pm$400~\kms, which is lower than the velocities we find for \ion{Fe}{ii} features ($\sim$3\,000 -- 4\,000~\kms) at the same epoch. Likewise, if the feature around $\sim$7\,200~\AA\, were solely due to \ion{C}{ii}~$\lambda$7\,234, by $+28.5$\,d this would indicate a velocity of $\sim$0~\kms. While lower velocities of \ion{C}{ii} relative to \ion{Si}{ii} and \ion{Fe}{ii} have been suggested for a few SNe~Iax \citep{comp--obs--12z, tomasella--2016, tomasella--20}, this raises the possibility of misidentification. \cite{maguire--23} argue the potential carbon features in SN~2020udy may be contaminated by \ion{Fe}{ii}. Assuming a velocity of $\sim$3\,000 -- 4\,000~\kms, a number of \ion{Fe}{ii} transitions could be responsible for the absorption at $\sim$6\,560~\AA\, that we identify as \ion{C}{ii}~$\lambda$6\,580. In all cases however, these transitions are populated by much higher energy levels than the \ion{Fe}{ii}~$\lambda$6\,149 \& $\lambda$6\,247 features. Conversely, the tentative \ion{C}{ii}~$\lambda$7\,234 absorption could instead be due to \ion{Fe}{ii}~$\lambda$7\,321 at $\sim$3\,500~\kms\, as this feature shares the same electron configuration as \ion{Fe}{ii}~$\lambda$6\,149 \& $\lambda$6\,247. While this identification is perhaps more likely than \ion{C}{ii}~$\lambda$7\,234 at $\sim$0~\kms, some \ion{C}{ii} absorption may still be present and the lack of a convincing alternative identification for \ion{C}{ii}~$\lambda$6\,580 indicates that we cannot rule out the presence of at least some carbon absorption with low velocities at later times.

\par

The presence of features consistent with carbon in our earliest and latest spectra indicates that carbon may be found throughout most, or indeed the entirety, of the ejecta in SN~2025qe. Qualitatively, this is in agreement with predictions from pure deflagration models, in which unburned material is entrained by the turbulent flame as it propagates throughout the white dwarf \citep{niemeyer--96, reinecke--02a, reinecke--02b, fink-2014, lach--22--def}. We note that similar features (at slightly lower velocities) are also observed in the spectra of SN~2019muj, which was the subject of detailed spectroscopic analysis by \cite{barna--21}. Using \textsc{tardis} \citep{tardis} simulations, \cite{barna--21} find that the strengths of \ion{C}{ii} features are too strong in their maximum light spectra when a uniform abundance model is used and argue instead that excluding carbon below 6\,500~\kms produces more favourable agreement. From our spectra, although some features may be contaminated by \ion{Fe}{ii}, we cannot exclude carbon from the low velocity region of SN~2025qe. Assuming density profiles from the \cite{fink-2014} pure deflagration models, these velocities correspond to the inner $\sim$0.01 -- 0.06~$M_{\odot}$ of ejecta. Together with the presence of carbon in our earliest spectra (i.e. the outermost regions of the ejecta), this would indicate a potentially well-mixed structure. Previous studies have shown that while carbon is predicted throughout the ejecta of pure deflagration models, the features are indeed stronger than observed in SNe~Iax \citep{kromer-13}, with mass fractions ranging from $\sim$30 -- 60 per cent \citep{jordan--12, fink-2014, lach--22--def}. The initial composition and ignition conditions of the progenitor white dwarf clearly can have an impact on the resulting ejecta. It is unclear however whether these factors are significant enough to produce the apparent reduction in carbon in the ejecta required by observations of SNe~Iax.

\subsection{Rise times}
\label{sect:rise_times}

Studies of individual SNe~Iax have highlighted their short rise times \citep{phillips--07, foley--09, obs--07qd, srivastav--22}. \cite{foley--13} note that the rise times of SNe~Iax range from $\sim$10\,d to $\textgreater$20\,d and may be systematically shorter than normal SNe~Ia. In agreement with this, our sample of 14 SNe~Iax observed by ATLAS, GOTO, and ZTF includes rise times ranging from 9.7\,d -- 26.2\,d across various filters. Furthermore, also in agreement with previous work, we find some evidence in favour of a correlation between rise time and peak absolute magnitude, with brighter SNe~Iax typically taking longer to reach maximum light. Normal SNe~Ia do not appear to show a similar correlation.

\par

Analytical modelling of the bolometric light curves of SNe~Iax indicates a lower ejecta mass compared with normal SNe~Ia \citep{mccully--14, srivastav--22}, but the differences in rise times among different bands highlights there could be significant opacity effects also playing a role. Our results show that redder bands have longer rise times compared with bluer bands for SNe~Iax. This is consistent with detailed studies of individual SNe~Iax (e.g. \citealt{02cx--orig, phillips--07, comp--obs--12z, maguire--23}) and most other types of SNe, but in contrast to normal SNe~Ia, which generally reach maximum first in the redder NIR ($IJHK$-) bands and at similar times in the optical $VR$ bands a few days later (e.g. \citealt{contardo--00, pastorello--07, pereira--13, cartier--15}). Of the 14 SNe~Iax included in our final sample analysis (Sect.~\ref{sect:early_lcs}), 9 have rise time measurements in at least two bands. The time between ZTF $g$- and $r$-band maxima range from $\sim$3 -- 5\,d. SN~2021jun is the only SN~Iax for which we measure a ZTF $i$-band rise time and we find it takes $\sim$1 week longer to reach maximum than the ZTF $g$-band (14\,d cf. 21\,d).

\par

We note that our rise time measurements are from the moment of first light (i.e. when the first photons escape the SN) to maximum and therefore do not include a dark phase \citep{piro-nakar-2013}. The length of the dark phase is proportional to the distribution of $^{56}$Ni within the ejecta, with more mixed ejecta having shorter dark phases \citep{piro-nakar-2014, piro-16, magee--20}. Therefore if SNe~Iax contain well-mixed ejecta, the rise times measured in this work will be relatively unaffected, but the rise times of normal SNe~Ia could be systematically longer by up to a few days if their ejecta is more stratified. This would likely further increase the differences in the rise time distributions of SNe~Iax and normal SNe~Ia. Verifying this speculation however requires detailed radiative transfer modelling of each SN to determine the true time of explosion, but this is beyond the scope of the work presented here.

\par

The rise times predicted by various pure deflagration \citep{fink-2014, kromer-15, lach--22--def} and pulsationally assisted gravitationally confined detonation (PGCD) models \citep{lach--22--def} are also given in Fig.~\ref{fig:abs_mag_vs_rise_time}. The rise times of both explosion scenarios are generally faster than observed among SNe~Iax, although in the case of pure deflagrations the inclusion of the bound remnant may improve agreement with observations \citep{callan--24}. Pure deflagration models follow the same general trend observed in SNe~Iax of longer rise times in redder bands and lack a prominent secondary maximum. The light curves predicted by PGCD models show a more complicated morphology reflecting their more complicated ejecta structure and as such do not necessarily show the same trend \citep{lach--22--pdd}. Pure deflagration models also show a strong correlation between peak absolute magnitude and rise time (Fig.~\ref{fig:abs_mag_vs_rise_time}), which qualitatively matches the observed SNe~Iax sample. Conversely, the PGCD models generally show a trend only in the bluest (ZTF $g$) and reddest (ZTF $i$) bands, while all other bands do not show a strong correlation and instead generally predict similar rise times. 

\par

\cite{kashyap--18} show that the bolometric rise time for their white dwarf merger model is $\sim$6\,d, which is shorter than the rise times we find in our SNe~Iax sample. Although not included in their light curve calculation, some amount of fallback material may also contribute to the luminosity and could lead to longer rise times \citep{shen--17}. \cite{kashyap--18} speculate that larger oxygen-neon primaries could also lead to higher ejecta and $^{56}$Ni masses. This could produce peak luminosities and rise times more similar to fainter members of the class such as SN~2024vjm (Fig.~\ref{fig:abs_mag_vs_rise_time}). We encourage further exploration of this merger scenario to determine the range of possible light curves and whether they are consistent with trends observed among SNe~Iax.

\subsection{Rise indices}
\label{sect:rise_indices}

Studies of individual SNe~Iax have typically found rise indices ranging from $\sim$1.0 -- 1.4 \citep{miller--18, maguire--23, hoogendam--25}. Consistent with this, our light curve fits show similar shallow rises and indicate that the rise indices of SNe~Iax are systematically lower than in normal SNe~Ia, with mean weighted rise indices of $\sim$1.4 -- 1.5 in the ZTF $g$- and $r$-bands compared to $\sim$2 for SNe~Ia.

\par

In addition to producing a shorter dark phase, an extended $^{56}$Ni distribution also produces a shallower rise towards maximum light (i.e. lower $\alpha$; \citealt{piro-nakar-2014}). \cite{magee--18} present radiative transfer models of different $^{56}$Ni distributions, with those containing $^{56}$Ni throughout the ejecta showing an average $B$-band rise index of $\sim$1.6 while the most compact $^{56}$Ni distributions show an average of $\sim$2.8. The $^{56}$Ni mass fraction for the most extended models presented by \cite{magee--18} decreases $\sim$90 per cent from the inner to outer ejecta. This is more extreme than the change in the (angle-averaged) $^{56}$Ni distributions predicted by pure deflagration models, which typically only vary by up to a few tens of per cent throughout the ejecta. Therefore, pure deflagrations and more vigorously mixed models may show even lower rise indices, similar to those observed among SNe~Iax.

\par

\cite{noebauer--17} calculate synthetic $UBVR$ light curves for a handful of specific explosion models from 10$^4$~s after explosion up to 10 days later. These models include pure deflagrations (N5def and N1600Cdef; \citealt{fink-2014}), the delayed detonation of a Chandrasekhar-mass white dwarf (N100; \citealt{seitenzahl--13}), the pure detonation of a 1.06~M$_{\odot}$ white dwarf \citep{sim--10}, and the violent merger of 1.1~M$_{\odot}$ and 0.9~M$_{\odot}$ carbon-oxygen white dwarfs \citep{pakmor-2012}. We apply the method outlined in Sect.~\ref{sect:analysis_lc} to measure the rise indices of these models, but we note that they do not extend up to maximum light and therefore we fit only up to half the peak flux available (i.e. half the flux at 10\,d post-explosion). For the pure deflagration models this is close to the time of maximum, but significantly shorter for all other models. As such, the rise indices measured are not directly comparable to our observed SNe~Iax sample although they are nevertheless useful reference points. For the N5def and N1600Cdef pure deflagration models, we find $B$-band rise indices of $\sim$1.4. All other scenarios show rise indices of $\geq$1.6. Both deflagration models show dark phases of $\leq$0.1\,d, while the other scenarios have dark phases of $\gtrsim$1\,d. The low rise indices predicted by pure deflagration models are qualitatively consistent with our findings for SNe~Iax. It is unclear however whether this scenario can produce the full range of rise indices measured here or indeed the nearly linear rises observed among some SNe. Indeed, the N5def model shows a nearly uniform angle-averaged $^{56}$Ni distribution and hence its rise index is unlikely to be reduced further by additional mixing. The viewing-angle dependence of the rise index has yet to be fully explored and therefore it is unclear whether this could account for even lower rise indices or indeed the level of variation expected. Additional models covering the earliest moments after explosion for a range of deflagration strengths are needed to fully test the viability of this scenario in reproducing the early light curves of SNe~Iax.

%

\section{Conclusions}
\label{sect:conclusions}
In this study, we presented photometric and spectroscopic observations of two new members of the SN~Iax class, SN~2024bfu and SN~2025qe. Both SNe were discovered shortly after explosion and our dataset for SN~2025qe in particular includes some of the earliest spectroscopic observations of any SN~Iax, beginning only 1.8\,d after the estimated time of first light. 

\par

Spectra of SN~2025qe show features consistent with carbon absorption throughout its evolution. From our earliest spectra, we identified features consistent with \ion{C}{iii}~$\lambda$4\,647 moving at similar velocities to those of \ion{Si}{iii} and \ion{Fe}{iii}. All spectra show tentative signs of \ion{C}{ii}~$\lambda$6\,580. Approximately one month after maximum light, the tentative \ion{C}{ii}~$\lambda$6\,580 feature may be contaminated partially or fully by \ion{Fe}{ii} absorption, but we are unable to identify a plausible alternative identification. \ion{C}{ii}~$\lambda$7\,234 has also been suggested for some SNe~Iax and we identified a similar feature in our later spectra of SN~2025qe. Based on the low inferred velocities however, we proposed an alternative identification of \ion{Fe}{ii}~$\lambda$7\,321, which would imply similar velocities to other \ion{Fe}{ii} features. The presence of carbon features throughout the spectral evolution would indicate that carbon cannot be excluded from the inner ejecta of SN~2025qe and therefore some amount of unburned material is well-mixed throughout.

\par

Both SN~2024bfu and SN~2025qe were observed by multiple all-sky surveys (ATLAS, GOTO, or ZTF) around explosion. As such, we were able to place tight constraints on their early light curves, including the epoch of first light and rise indices in various bands. We found that both SNe showed relatively shallow rises up to maximum light. Inspired by these constraints and the possibilities afforded by combining multiple surveys, we gathered a sample of 14 SNe~Iax that were publicly classified on the TNS and obtained their forced photometry light curves from these surveys. For comparison purposes we also gathered a sample of 87 normal SNe~Ia. Consistent with previous work, we found that SNe~Iax may show systematically shorter rise times than normal SNe~Ia. In addition, we also found that SNe~Iax may show systematically lower rise indices (shallower rises to maximum). These differences are most pronounced in the bluer bands, but more work and larger samples may show whether or not this is statistically significant. Our results generally support previous suggestions in the literature of a correlation between peak absolute magnitude and rise time, but we do not see strong evidence in favour of a correlation between rise time and rise index, which has also been claimed in the literature. This likely arises from the smaller sample size and larger uncertainties used in this work. Although combining multiple surveys, our results were still limited by small sample sizes and relatively large uncertainties. In particular rise index measurements are specific to each band and therefore do not benefit from the combination of multiple surveys observing with their own unique filters. Higher cadence surveys will enable tighter constraints on the rise index in particular, which will help to assess the statistical significance of trends found here. Alternatively, consistent filter sets across multiple surveys will also naturally lead to better-sampled light curves and tighter constraints. 

\par

The low rise indices observed among SNe~Iax, and the longer rise times in redder bands, are consistent with expectations for a well-mixed ejecta as more extended distributions of $^{56}$Ni typically result in shallower rises towards maximum light. These properties are also consistent with predictions from the pure deflagration scenario, but in contrast to some pulsationally-assisted gravitationally confined detonations. Therefore, our analysis generally supports the conclusion that significant mixing, similar to that predicted by pure deflagration models, is imparted on the ejecta of SNe~Iax. It is unclear however, whether this scenario can explain the variation in rise indices observed or indeed the nearly linear rises measured for some SNe~Iax. Merger scenarios have also been proposed as a viable mechanism for SNe~Iax however the level of mixing predicted by such mergers is currently unclear. We encourage further exploration of mergers to determine whether this scenario can reproduce the observed photometric and spectroscopic features of SNe~Iax and the trends identified here.

%

\section*{Acknowledgements}

We thank the referee for their comments, which helped to improve the clarity of this manuscript. 
MRM and TLK acknowledge Warwick Astrophysics prize post-doctoral fellowships made possible thanks to a generous philanthropic donation. 
JDL and MP acknowledge support from a UK Research and Innovation Future Leaders Fellowship (MR/T020784/1). 
BG acknowledges the UKRI's STFC studentship grant funding, project reference ST/X508871/1.
CJP acknowledges financial support from grant PRE2021-096988 funded by AEI 10.13039/501100011033 and ESF Investing in your future. 
LG acknowledges financial support from AGAUR, CSIC, MCIN and AEI 10.13039/501100011033 under projects PID2023-151307NB-I00, PIE 20215AT016, CEX2020-001058-M, ILINK23001, COOPB2304, and 2021-SGR-01270. 
SGG acknowledges support from the ESO Scientiﬁc Visitor Programme. 
LK acknowledges support for an Early Career Fellowship from the Leverhulme Trust through grant ECF-2024-054 and the Isaac Newton Trust through grant 24.08(w). 
GL was supported by a research grant (VIL60862) from VILLUM FONDEN. 
SM is funded by Leverhulme Trust grant RPG-2023-240.
TEMB is funded by Horizon Europe ERC grant no. 101125877. 
RWW and TB acknowledge financial support from Science and Technology Facilities Council (STFC, grant number ST/X001075/1). 
This work was funded by ANID, Millennium Science Initiative, ICN12\_009. 
We derive posterior probability distributions and the Bayesian evidence with the nested sampling Monte Carlo algorithm MLFriends (Buchner, 2014; 2019) using the UltraNest\footnote{\url{https://johannesbuchner.github.io/UltraNest/}} package (Buchner 2021). 
The Gravitational-wave Optical Transient Observer (GOTO) project acknowledges support from the Science and Technology Facilities Council (STFC, grant numbers ST/T007184/1, ST/T003103/1, ST/T000406/1, ST/X001121/1 and ST/Z000165/1) and the GOTO consortium institutions; University of Warwick; Monash University; University of Sheffield; University of Leicester; Armagh Observatory \& Planetarium; the National Astronomical Research Institute of Thailand (NARIT); University of Manchester; Instituto de Astrofísica de Canarias (IAC); University of Portsmouth; University of Turku.
The Liverpool Telescope is operated on the island of La Palma by Liverpool John Moores University in the Spanish Observatorio del Roque de los Muchachos of the Instituto de Astrofisica de Canarias with financial support from the UK Science and Technology Facilities Council. 
The Isaac Newton Telescope is operated on the island of La Palma by the Isaac Newton Group of Telescopes in the Spanish Observatorio del Roque de los Muchachos of the Instituto de Astrofísica de Canarias. 
The pt5m telescope is supported by the Isaac Newton Group of Telescopes in La Palma.
This article includes observations made by the Two-meter Twin Telescope (TTT) in the Teide Observatory of the IAC, that Light Bridges operates in the Island of Tenerife, Canary Islands (Spain). The Observing Time Rights used for this research were provided by Light Bridges, SL. 
IDS spectroscopy was obtained as part of 2024P07 on the Isaac Newton Telescope. The Isaac Newton Telescope is operated on the island of La Palma by the Isaac Newton Group of Telescopes in the Spanish Observatorio del Roque de los Muchachos of the Instituto de Astrofísica de Canarias.
Based on observations collected at the European Organisation for Astronomical Research in the Southern Hemisphere, Chile, as part of ePESSTO+ (the advanced Public ESO Spectroscopic Survey for Transient Objects Survey – PI: Inserra) ePESSTO+ observations were obtained under ESO program ID 112.25JQ. 
Based on observations collected at Centro Astronómico Hispano en Andalucía (CAHA) at Calar Alto, proposal 25A-2.2-003, operated jointly by Junta de Andalucía and Consejo Superior de Investigaciones Científicas (IAA-CSIC). 
The ZTF forced-photometry service was funded under the Heising-Simons Foundation grant\#12540303 (PI: Graham). 
This work has made use of data from the Asteroid Terrestrial-impact Last Alert System (ATLAS) project. The Asteroid Terrestrial-impact Last Alert System (ATLAS) project is primarily funded to search for near earth asteroids through NASA grants NN12AR55G, 80NSSC18K0284, and 80NSSC18K1575; byproducts of the NEO search include images and catalogs from the survey area. This work was partially funded by Kepler/K2 grant J1944/80NSSC19K0112 and HST GO-15889, and STFC grants ST/T000198/1 and ST/S006109/1. The ATLAS science products have been made possible through the contributions of the University of Hawaii Institute for Astronomy, the Queen’s University Belfast, the Space Telescope Science Institute, the South African Astronomical Observatory, and The Millennium Institute of Astrophysics (MAS), Chile. 
The Pan-STARRS1 Surveys (PS1) and the PS1 public science archive have been made possible through contributions by the Institute for Astronomy, the University of Hawaii, the Pan-STARRS Project Office, the Max-Planck Society and its participating institutes, the Max Planck Institute for Astronomy, Heidelberg and the Max Planck Institute for Extraterrestrial Physics, Garching, The Johns Hopkins University, Durham University, the University of Edinburgh, the Queen's University Belfast, the Harvard-Smithsonian Center for Astrophysics, the Las Cumbres Observatory Global Telescope Network Incorporated, the National Central University of Taiwan, the Space Telescope Science Institute, the National Aeronautics and Space Administration under Grant No. NNX08AR22G issued through the Planetary Science Division of the NASA Science Mission Directorate, the National Science Foundation Grant No. AST–1238877, the University of Maryland, Eotvos Lorand University (ELTE), the Los Alamos National Laboratory, and the Gordon and Betty Moore Foundation. 
This work has made use of data from the European Space Agency (ESA) mission {\it Gaia} (\url{https://www.cosmos.esa.int/gaia}), processed by the {\it Gaia} Data Processing and Analysis Consortium (DPAC, \url{https://www.cosmos.esa.int/web/gaia/dpac/consortium}). Funding for the DPAC has been provided by national institutions, in particular the institutions participating in the {\it Gaia} Multilateral Agreement. 

\section*{Data Availability}
Spectra are publicly available on WISeREP.



\bibliographystyle{mnras}
\bibliography{Magee}




\appendix

\section{Filters}
\label{appendix_filters}
\begin{figure}
\centering
\includegraphics[width=\columnwidth]{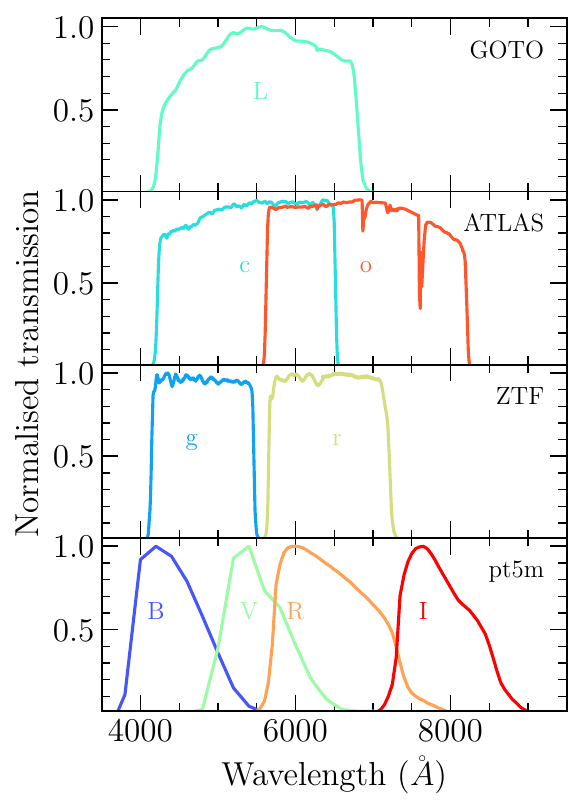}
\caption{Normalised transmission curves for filters used in this work.}
\label{fig:filters}
\centering
\end{figure}

Figure~\ref{fig:filters} shows the filters used throughout this work. The GOTO $L$-band is a broad filter extending from $\sim$4\,000 -- 7\,000~\AA\, and covers a similar wavelength range as the ZTF $g$- and $r$-bands. To demonstrate the properties of this filter we calculate light curves for both normal SNe~Ia and SNe~Iax in the $L$-band and more standard $BV$ bands. In both cases we find that the GOTO $L$-band is most similar to the $V$-band.
\par
We use SALT2 \citep{guy--07} to generate synthetic light curves of normal SNe~Ia for a range of $x_1$ values ($-$2 -- 2). We find only minor changes in both the time and magnitude of peak. In the GOTO $L$-band, the SALT2 model produces a peak approximately 1.3 -- 1.6 days after $B$-band maximum and $\sim$0.5 days before $V$-band maximum. The peak magnitude in the GOTO $L$-band is between 0 -- 0.03~mag fainter than in the $B$- or $V$-bands.
\par
For SNe~Iax we use the PLAsTiCC templates presented by \cite{kessler--19} to generate synthetic light curves. Again, we find only minor changes in both the time and magnitude of peak, although these are more pronounced than in normal SNe~Ia. The PLAsTiCC templates typically peak in the GOTO $L$-band 2 -- 3 days after $B$-band maximum and 0.7 -- 1 day before $V$-band maximum. The peak magnitudes show only minor variations, with the GOTO $L$-band being $\sim$0.05~mag brighter/fainter than the $B$-/$V$-bands, respectively.

\section{Photometric logs}
\label{appendix_phot}

\onecolumn

\begin{center}
\footnotesize{
\begin{longtable}{lcrrrr}
\caption{Survey photometry of SN~2024bfu.} \\
\hline
\multicolumn{1}{c}{\textbf{Date}}	 &	\multicolumn{1}{c}{\textbf{MJD}} &  \multicolumn{1}{c}{\textbf{Phase}$^{a}$}   & \multicolumn{1}{c}{\textbf{$L$}}  &	\multicolumn{1}{c}{\textbf{$c$}} &	\multicolumn{1}{c}{\textbf{$o$}} \\ 
\multicolumn{1}{l}{}	             &	\multicolumn{1}{c}{}     &  \multicolumn{1}{c}{\textbf{(days)}}	               & \multicolumn{1}{c}{\textbf{($\mu$Jy)}}	   &	\multicolumn{1}{c}{\textbf{($\mu$Jy)}} &  \multicolumn{1}{c}{\textbf{($\mu$Jy)}} \\ \hline\hline
\label{tab:24bfu_phot}
\endfirsthead
\multicolumn{6}{c}%
{{\tablename\ \thetable{} -- continued from previous page}} \\
\hline
\multicolumn{1}{c}{\textbf{Date}}	 &	\multicolumn{1}{c}{\textbf{MJD}} &  \multicolumn{1}{c}{\textbf{Phase}$^{a}$}   & \multicolumn{1}{c}{\textbf{$L$}}  &	\multicolumn{1}{c}{\textbf{$c$}} &	\multicolumn{1}{c}{\textbf{$o$}} \\ 
\multicolumn{1}{l}{}	             &	\multicolumn{1}{c}{}     &  \multicolumn{1}{c}{\textbf{(days)}}	               & \multicolumn{1}{c}{\textbf{($\mu$Jy)}}	   &	\multicolumn{1}{c}{\textbf{($\mu$Jy)}} &  \multicolumn{1}{c}{\textbf{($\mu$Jy)}} \\ \hline\hline
\hline\hline
\endhead

\hline \multicolumn{6}{|r|}{{Continued on next page}} \\ \hline
\endfoot

\endlastfoot
2024 01 21 & 60\,330.24 & $-$16.92 		& $\cdots$ 			& $\cdots$ 			& 10.25$\pm$8.72 \\
2024 01 21 & 60\,330.58 & $-$16.59 		& 1.15$\pm$7.48 	& $\cdots$ 			& $\cdots$ \\
2024 01 22 & 60\,331.54 & $-$15.67 		& 11.92$\pm$10.44 	& $\cdots$ 			& $\cdots$ \\
2024 01 23 & 60\,332.54 & $-$14.70 		& -14.83$\pm$12.50 	& $\cdots$ 			& $\cdots$ \\
2024 01 24 & 60\,333.24 & $-$14.03 		& $\cdots$ 			& $\cdots$ 			& -15.50$\pm$19.94 \\
2024 01 24 & 60\,333.54 & $-$13.74 		& 10.80$\pm$9.06 	& $\cdots$ 			& $\cdots$ \\
2024 01 25 & 60\,334.25 & $-$13.05 		& $\cdots$ 			& $\cdots$ 			& 38.75$\pm$19.75 \\
2024 01 25 & 60\,334.56 & $-$12.75 		& 42.88$\pm$10.46 	& $\cdots$ 			& $\cdots$ \\
2024 01 27 & 60\,336.54 & $-$10.84 		& 155.43$\pm$120.74 & $\cdots$ 			& $\cdots$ \\
2024 01 28 & 60\,337.26 & $-$10.14 		& $\cdots$ 			& $\cdots$ 			& 96.50$\pm$12.97 \\
2024 01 28 & 60\,337.53 & $-$9.88 		& 84.61$\pm$15.25 	& $\cdots$ 			& $\cdots$ \\
2024 01 29 & 60\,338.22 & $-$9.22 		& $\cdots$ 			& $\cdots$ 			& 120.00$\pm$12.19 \\
2024 01 31 & 60\,340.01 & $-$7.49 		& $\cdots$ 			& $\cdots$ 			& 155.00$\pm$11.98 \\
2024 01 31 & 60\,340.55 & $-$6.97 		& 151.71$\pm$8.82 	& $\cdots$ 			& $\cdots$ \\
2024 01 31 & 60\,340.56 & $-$6.96 		& 175.83$\pm$9.89 	& $\cdots$ 			& $\cdots$ \\
2024 02 01 & 60\,341.21 & $-$6.33 		& $\cdots$ 			& $\cdots$ 			& 173.33$\pm$10.12 \\
2024 02 01 & 60\,341.49 & $-$6.06 		& 150.35$\pm$4.56 	& $\cdots$ 			& $\cdots$ \\
2024 02 01 & 60\,341.57 & $-$5.98 		& 164.81$\pm$5.66 	& $\cdots$ 			& $\cdots$ \\
2024 02 01 & 60\,341.66 & $-$5.90 		& 160.49$\pm$10.21 	& $\cdots$ 			& $\cdots$ \\
2024 02 02 & 60\,342.21 & $-$5.37 		& $\cdots$ 			& $\cdots$ 			& 189.50$\pm$8.04 \\
2024 02 02 & 60\,342.47 & $-$5.12 		& 169.85$\pm$6.90 	& $\cdots$ 			& $\cdots$ \\
2024 02 02 & 60\,342.58 & $-$5.01 		& 172.45$\pm$9.09 	& $\cdots$ 			& $\cdots$ \\
2024 02 02 & 60\,342.64 & $-$4.95 		& 173.20$\pm$14.42 	& $\cdots$ 			& $\cdots$ \\
2024 02 03 & 60\,343.11 & $-$4.50 		& $\cdots$ 			& $\cdots$ 			& 191.67$\pm$16.32 \\
2024 02 03 & 60\,343.47 & $-$4.15 		& 176.70$\pm$6.33 	& $\cdots$ 			& $\cdots$ \\
2024 02 05 & 60\,345.22 & $-$2.46 		& $\cdots$ 			& $\cdots$ 			& 215.75$\pm$31.19 \\
2024 02 06 & 60\,346.20 & $-$1.52 		& $\cdots$ 			& $\cdots$ 			& 238.50$\pm$14.61 \\
2024 02 07 & 60\,347.94 & 0.16 			& $\cdots$ 			& $\cdots$ 			& 234.00$\pm$9.39 \\
2024 02 09 & 60\,349.19 & 1.37 			& $\cdots$ 			& $\cdots$ 			& 236.50$\pm$7.98 \\
2024 02 09 & 60\,349.47 & 1.64 			& 191.25$\pm$8.66 	& $\cdots$ 			& $\cdots$ \\
2024 02 10 & 60\,350.94 & 3.06 			& $\cdots$ 			& $\cdots$ 			& 233.25$\pm$6.98 \\
2024 02 11 & 60\,351.92 & 4.01 			& $\cdots$ 			& $\cdots$ 			& 249.25$\pm$6.68 \\
2024 02 12 & 60\,352.46 & 4.53 			& 178.66$\pm$8.14 	& $\cdots$ 			& $\cdots$ \\
2024 02 12 & 60\,352.63 & 4.69 			& 178.46$\pm$5.20 	& $\cdots$ 			& $\cdots$ \\
2024 02 13 & 60\,353.20 & 5.24 			& $\cdots$ 			& $\cdots$ 			& 286.00$\pm$8.73 \\
2024 02 13 & 60\,353.45 & 5.48 			& 167.08$\pm$7.37 	& $\cdots$ 			& $\cdots$ \\
2024 02 13 & 60\,353.62 & 5.65 			& 167.64$\pm$7.06 	& $\cdots$ 			& $\cdots$ \\
2024 02 14 & 60\,354.18 & 6.19 			& $\cdots$ 			& $\cdots$ 			& 272.50$\pm$7.50 \\
2024 02 17 & 60\,357.10 & 9.01 			& $\cdots$ 			& $\cdots$ 			& 272.25$\pm$9.60 \\
2024 02 18 & 60\,358.55 & 10.41 		& $\cdots$ 			& $\cdots$ 			& 229.00$\pm$5.59 \\
2024 02 19 & 60\,359.92 & 11.73 		& $\cdots$ 			& $\cdots$ 			& 217.25$\pm$8.59 \\
2024 02 21 & 60\,361.17 & 12.93 		& $\cdots$ 			& $\cdots$ 			& 218.00$\pm$14.30 \\
2024 02 22 & 60\,362.57 & 14.29 		& $\cdots$ 			& $\cdots$ 			& 208.17$\pm$15.81 \\
2024 02 23 & 60\,363.68 & 15.36 		& $\cdots$ 			& $\cdots$ 			& 169.75$\pm$17.22 \\
2024 02 25 & 60\,365.19 & 16.81 		& $\cdots$ 			& $\cdots$ 			& 121.00$\pm$28.72 \\
2024 02 26 & 60\,366.57 & 18.15 		& $\cdots$ 			& $\cdots$ 			& 142.88$\pm$10.51 \\
2024 02 27 & 60\,367.95 & 19.48 		& $\cdots$ 			& $\cdots$ 			& 143.25$\pm$11.84 \\
2024 02 29 & 60\,369.27 & 20.75 		& $\cdots$ 			& $\cdots$ 			& 129.25$\pm$9.41 \\
2024 03 01 & 60\,370.93 & 22.36 		& $\cdots$ 			& $\cdots$ 			& 134.00$\pm$7.60 \\
2024 03 02 & 60\,371.93 & 23.32 		& $\cdots$ 			& $\cdots$ 			& 122.75$\pm$8.70 \\
2024 03 05 & 60\,374.89 & 26.18 		& $\cdots$ 			& 41.50$\pm$4.84 	& $\cdots$ \\
2024 03 06 & 60\,375.93 & 27.18 		& $\cdots$ 			& 47.00$\pm$5.80 	& $\cdots$ \\
2024 03 09 & 60\,378.14 & 29.31 		& $\cdots$ 			& $\cdots$ 			& 78.75$\pm$7.07 \\
2024 03 09 & 60\,378.89 & 30.04 		& $\cdots$ 			& 46.75$\pm$4.73 	& $\cdots$ \\
2024 03 10 & 60\,379.86 & 30.97 		& $\cdots$ 			& 44.00$\pm$5.35 	& $\cdots$ \\
2024 03 12 & 60\,381.14 & 32.21 		& $\cdots$ 			& $\cdots$ 			& 89.75$\pm$6.79 \\
2024 03 12 & 60\,381.47 & 32.53 		& 40.70$\pm$3.58 	& $\cdots$ 			& $\cdots$ \\
2024 03 13 & 60\,382.13 & 33.17 		& $\cdots$ 			& $\cdots$ 			& 105.75$\pm$8.74 \\
2024 03 13 & 60\,382.46 & 33.48 		& 28.50$\pm$4.11 	& $\cdots$ 			& $\cdots$ \\
2024 03 13 & 60\,382.90 & 33.91 		& $\cdots$ 			& 43.75$\pm$4.87 	& $\cdots$ \\
2024 03 14 & 60\,383.89 & 34.86 		& $\cdots$ 			& 34.00$\pm$5.52 	& $\cdots$ \\
2024 03 15 & 60\,384.86 & 35.80 		& $\cdots$ 			& 34.33$\pm$5.51 	& $\cdots$ \\
2024 03 17 & 60\,386.40 & 37.29 		& $\cdots$ 			& $\cdots$ 			& 66.86$\pm$12.90 \\
2024 03 18 & 60\,387.88 & 38.72 		& $\cdots$ 			& $\cdots$ 			& 86.25$\pm$7.62 \\
2024 03 19 & 60\,388.45 & 39.27 		& 29.69$\pm$5.55 	& $\cdots$ 			& $\cdots$ \\
2024 03 20 & 60\,389.13 & 39.92 		& $\cdots$ 			& $\cdots$ 			& 74.00$\pm$26.21 \\
2024 03 21 & 60\,390.48 & 41.23 		& $\cdots$ 			& $\cdots$ 			& 65.12$\pm$9.05 \\
2024 03 22 & 60\,391.44 & 42.15 		& 32.27$\pm$7.25 	& $\cdots$ 			& $\cdots$ \\
2024 03 25 & 60\,394.44 & 45.05 		& 44.04$\pm$15.90 	& $\cdots$ 			& $\cdots$ \\
2024 03 25 & 60\,394.61 & 45.21 		& $\cdots$ 			& $\cdots$ 			& 44.00$\pm$11.89 \\
2024 03 26 & 60\,395.43 & 46.00 		& 39.55$\pm$8.22 	& $\cdots$ 			& $\cdots$ \\
2024 03 26 & 60\,395.90 & 46.46 		& $\cdots$ 			& $\cdots$ 			& 27.67$\pm$18.28 \\
2024 03 27 & 60\,396.43 & 46.97 		& 27.89$\pm$13.57 	& $\cdots$ 			& $\cdots$ \\
2024 03 29 & 60\,398.46 & 48.93 		& $\cdots$ 			& $\cdots$ 			& 57.57$\pm$8.73 \\
2024 03 30 & 60\,399.91 & 50.33 		& $\cdots$ 			& $\cdots$ 			& 71.50$\pm$21.82 \\
\hline\hline
\multicolumn{6}{l}{$^{a}$ Phases are given relative to GOTO $L$-band maximum, MJD = 60\,347.77.} \\
\end{longtable}
}
\end{center}

\begin{center}
\footnotesize{
\begin{longtable}{lcrrrr}
\caption{LT photometry of SN~2024bfu.}  \\
\hline
\multicolumn{1}{c}{\textbf{Date}}	 &	\multicolumn{1}{c}{\textbf{MJD}} &  \multicolumn{1}{c}{\textbf{Phase}$^{a}$}  &	\multicolumn{1}{c}{\textbf{$g$}} &	\multicolumn{1}{c}{\textbf{$r$}} &	\multicolumn{1}{c}{\textbf{$i$}}   \\ 
\multicolumn{1}{l}{}	             &	\multicolumn{1}{c}{}     &  \multicolumn{1}{c}{\textbf{(days)}}	                &  \multicolumn{1}{c}{\textbf{($\mu$Jy)}} &  \multicolumn{1}{c}{\textbf{($\mu$Jy)}} &  \multicolumn{1}{c}{\textbf{($\mu$Jy)}}  \\ \hline\hline
\label{tab:24bfu_lt_phot}
\endfirsthead
\multicolumn{6}{c}%
{{\tablename\ \thetable{} -- continued from previous page}} \\
\hline
\multicolumn{1}{c}{\textbf{Date}}	 &	\multicolumn{1}{c}{\textbf{MJD}} &  \multicolumn{1}{c}{\textbf{Phase}$^{a}$}  &	\multicolumn{1}{c}{\textbf{$g$}} &	\multicolumn{1}{c}{\textbf{$r$}} &	\multicolumn{1}{c}{\textbf{$i$}}   \\ 
\multicolumn{1}{l}{}	             &	\multicolumn{1}{c}{}     &  \multicolumn{1}{c}{\textbf{(days)}}	                &  \multicolumn{1}{c}{\textbf{($\mu$Jy)}} &  \multicolumn{1}{c}{\textbf{($\mu$Jy)}} &  \multicolumn{1}{c}{\textbf{($\mu$Jy)}}  \\ \hline\hline
\endhead

\hline \multicolumn{6}{|r|}{{Continued on next page}} \\ \hline
\endfoot

\endlastfoot

2024 02 06  & 60\,346.96 & $-$0.78 &  $\cdots$          & $\cdots$          & 205.68$\pm$4.74 \\
2024 02 11  & 60\,351.94 & 4.03    &  161.14$\pm$2.52   & 242.55$\pm$3.13   & 228.88$\pm$2.32 \\
2024 02 13  & 60\,353.99 & 6.00    &  139.83$\pm$2.19   & 236.37$\pm$3.05   & 238.34$\pm$2.41 \\
2024 02 16  & 60\,356.91 & 8.82    &  112.93$\pm$12.27  & 227.61$\pm$12.58  & 240.10$\pm$8.40 \\
2024 02 18  & 60\,358.93 & 10.77    &  83.41$\pm$14.21   & 198.61$\pm$12.44  & 237.03$\pm$7.64 \\
\hline\hline
\multicolumn{6}{l}{$^{a}$ Phases are given relative to GOTO $L$-band maximum, MJD = 60\,347.77.} \\
\end{longtable}
}
\end{center}


\begin{center}
\footnotesize{
\begin{longtable}{lcrrrrrr}
\caption{Survey photometry of SN~2025qe.}  \\
\hline
\multicolumn{1}{c}{\textbf{Date}}	 &	\multicolumn{1}{c}{\textbf{MJD}} &  \multicolumn{1}{c}{\textbf{Phase}$^{a}$}   & 	\multicolumn{1}{c}{\textbf{$g$}} &	\multicolumn{1}{c}{\textbf{$c$}} &	\multicolumn{1}{c}{\textbf{$L$}} &	\multicolumn{1}{c}{\textbf{$r$}} &	\multicolumn{1}{c}{\textbf{$o$}}  \\ 
\multicolumn{1}{l}{}	             &	\multicolumn{1}{c}{}     &  \multicolumn{1}{c}{\textbf{(days)}}	               & \multicolumn{1}{c}{\textbf{($\mu$Jy)}}	   &	\multicolumn{1}{c}{\textbf{($\mu$Jy)}} &  \multicolumn{1}{c}{\textbf{($\mu$Jy)}} &  \multicolumn{1}{c}{\textbf{($\mu$Jy)}} &  \multicolumn{1}{c}{\textbf{($\mu$Jy)}}  \\ \hline\hline
\label{tab:25qe_survey_phot}
\endfirsthead
\multicolumn{8}{c}%
{{\tablename\ \thetable{} -- continued from previous page}} \\
\hline
\multicolumn{1}{c}{\textbf{Date}}	 &	\multicolumn{1}{c}{\textbf{MJD}} &  \multicolumn{1}{c}{\textbf{Phase}$^{a}$}   & 	\multicolumn{1}{c}{\textbf{$g$}} &	\multicolumn{1}{c}{\textbf{$c$}} &	\multicolumn{1}{c}{\textbf{$L$}} &	\multicolumn{1}{c}{\textbf{$r$}} &	\multicolumn{1}{c}{\textbf{$o$}} \\ 
\multicolumn{1}{l}{}	             &	\multicolumn{1}{c}{}     &  \multicolumn{1}{c}{\textbf{(days)}}	               & \multicolumn{1}{c}{\textbf{($\mu$Jy)}}	   &	\multicolumn{1}{c}{\textbf{($\mu$Jy)}} &  \multicolumn{1}{c}{\textbf{($\mu$Jy)}} &  \multicolumn{1}{c}{\textbf{($\mu$Jy)}} &  \multicolumn{1}{c}{\textbf{($\mu$Jy)}}  \\ \hline\hline
\endhead

\hline \multicolumn{8}{|r|}{{Continued on next page}} \\ \hline
\endfoot

\endlastfoot
2025 01 10 & 60\,685.46 &  $-$20.59 	& $\cdots$ 			& $\cdots$ 			& $\cdots$ 			& $\cdots$ 			& -23.75$\pm$21.94 \\
2025 01 11 & 60\,686.04 &  $-$20.01 	& $\cdots$ 			& $\cdots$ 			& 41.12$\pm$27.88 	& $\cdots$ 			& $\cdots$ \\
2025 01 12 & 60\,687.34 &  $-$18.72 	& $\cdots$ 			& $\cdots$ 			& $\cdots$ 			& -2.21$\pm$13.62 	& $\cdots$ \\
2025 01 12 & 60\,687.49 &  $-$18.57 	& $\cdots$ 			& $\cdots$ 			& $\cdots$ 			& $\cdots$ 			& 14.00$\pm$24.44 \\
2025 01 13 & 60\,688.02 &  $-$18.04 	& $\cdots$ 			& $\cdots$ 			& -44.34$\pm$30.93 	& $\cdots$ 			& $\cdots$ \\
2025 01 14 & 60\,689.30 &  $-$16.77 	& $\cdots$ 			& $\cdots$ 			& $\cdots$ 			& -29.32$\pm$19.64 	& $\cdots$ \\
2025 01 16 & 60\,691.01 &  $-$15.07 	& $\cdots$ 			& $\cdots$ 			& -18.00$\pm$31.94 	& $\cdots$ 			& $\cdots$ \\
2025 01 16 & 60\,691.38 &  $-$14.71 	& $\cdots$ 			& $\cdots$ 			& $\cdots$ 			& -6.28$\pm$6.31 	& $\cdots$ \\
2025 01 16 & 60\,691.51 &  $-$14.58 	& $\cdots$ 			& $\cdots$ 			& $\cdots$ 			& $\cdots$ 			& 0.43$\pm$9.57 \\
2025 01 18 & 60\,693.00 &  $-$13.10 	& $\cdots$ 			& $\cdots$ 			& 121.13$\pm$22.29 	& $\cdots$ 			& $\cdots$ \\
2025 01 18 & 60\,693.26 &  $-$12.84 	& 117.85$\pm$5.07 	& $\cdots$ 			& $\cdots$ 			& $\cdots$ 			& $\cdots$ \\
2025 01 18 & 60\,693.37 &  $-$12.73 	& $\cdots$ 			& $\cdots$ 			& $\cdots$ 			& 119.40$\pm$4.87 	& $\cdots$ \\
2025 01 18 & 60\,693.46 &  $-$12.64 	& $\cdots$ 			& $\cdots$ 			& $\cdots$ 			& $\cdots$ 			& 110.50$\pm$9.38 \\
2025 01 19 & 60\,694.94 &  $-$11.17 	& $\cdots$ 			& $\cdots$ 			& 333.49$\pm$12.13 	& $\cdots$ 			& $\cdots$ \\
2025 01 20 & 60\,695.20 &  $-$10.91 	& 373.93$\pm$4.62 	& $\cdots$ 			& $\cdots$ 			& $\cdots$ 			& $\cdots$ \\
2025 01 20 & 60\,695.37 &  $-$10.74 	& $\cdots$ 			& $\cdots$ 			& $\cdots$ 			& 359.28$\pm$4.83 	& $\cdots$ \\
2025 01 20 & 60\,695.47 &  $-$10.65 	& $\cdots$ 			& $\cdots$ 			& $\cdots$ 			& $\cdots$			& 333.75$\pm$10.14 \\
2025 01 22 & 60\,697.42 &  $-$8.71 		& $\cdots$ 			& $\cdots$ 			& $\cdots$ 			& $\cdots$			& 541.25$\pm$10.69 \\
2025 01 23 & 60\,698.14 &  $-$7.99 		& 670.99$\pm$10.58 	& $\cdots$ 			& $\cdots$ 			& $\cdots$			& $\cdots$ \\
2025 01 23 & 60\,698.33 &  $-$7.81 		& $\cdots$ 			& $\cdots$ 			& $\cdots$ 			& 662.86$\pm$7.20 	& $\cdots$ \\
2025 01 24 & 60\,699.45 &  $-$6.69 		& $\cdots$ 			& 819.50$\pm$6.33 	& $\cdots$ 			& $\cdots$ 			& $\cdots$ \\
2025 01 25 & 60\,700.15 &  $-$6.00 		& $\cdots$ 			& $\cdots$ 			& 824.46$\pm$18.42 	& $\cdots$ 			& $\cdots$ \\
2025 01 26 & 60\,701.98 &  $-$4.18 		& $\cdots$ 			& $\cdots$ 			& 940.97$\pm$13.19 	& $\cdots$ 			& $\cdots$ \\
2025 01 28 & 60\,703.98 &  $-$2.19 		& $\cdots$ 			& $\cdots$ 			& 1065.15$\pm$18.73 & $\cdots$ 			& $\cdots$ \\
2025 01 30 & 60\,705.30 &  $-$0.88 		& $\cdots$ 			& $\cdots$ 			& $\cdots$ 			& 1038.14$\pm$7.38 	& $\cdots$ \\
2025 02 02 & 60\,708.96 &  2.75 		& $\cdots$ 			& $\cdots$ 			& 990.32$\pm$14.55 	& $\cdots$ 			& $\cdots$ \\
2025 02 03 & 60\,709.44 &  3.23 		& $\cdots$ 			& $\cdots$ 			& $\cdots$ 			& $\cdots$ 			& 1043.25$\pm$9.39 \\
2025 02 04 & 60\,710.19 &  3.97 		& $\cdots$ 			& $\cdots$ 			& $\cdots$ 			& 1169.68$\pm$7.23 	& $\cdots$ \\
2025 02 04 & 60\,710.33 &  4.11 		& $\cdots$ 			& $\cdots$ 			& $\cdots$ 			& 1254.62$\pm$39.76 & $\cdots$ \\
2025 02 04 & 60\,710.42 &  4.20 		& $\cdots$ 			& $\cdots$ 			& $\cdots$ 			& $\cdots$ 			& 1048.00$\pm$18.00 \\
2025 02 04 & 60\,710.96 &  4.74 		& $\cdots$ 			& $\cdots$ 			& 907.80$\pm$17.98 	& $\cdots$ 			& $\cdots$ \\
2025 02 05 & 60\,711.60 &  5.37 		& $\cdots$ 			& $\cdots$ 			& $\cdots$ 			& $\cdots$ 			& 1058.25$\pm$10.89 \\
2025 02 06 & 60\,712.10 &  5.87 		& $\cdots$ 			& $\cdots$ 			& 848.45$\pm$21.62 	& $\cdots$ 			& $\cdots$ \\
2025 02 06 & 60\,712.88 &  6.64 		& $\cdots$ 			& $\cdots$ 			& 784.13$\pm$27.24 	& $\cdots$ 			& $\cdots$ \\
2025 02 07 & 60\,713.14 &  6.90 		& $\cdots$ 			& $\cdots$ 			& 781.12$\pm$20.42 	& $\cdots$ 			& $\cdots$ \\
2025 02 09 & 60\,715.44 &  9.19 		& $\cdots$ 			& $\cdots$ 			& $\cdots$ 			& $\cdots$ 			& 963.25$\pm$61.11 \\
2025 02 10 & 60\,716.38 &  10.12 		& 279.63$\pm$7.99 	& $\cdots$ 			& $\cdots$ 			& $\cdots$ 			& $\cdots$ \\
2025 02 11 & 60\,717.45 &  11.18 		& $\cdots$ 			& $\cdots$ 			& $\cdots$ 			& $\cdots$ 			& 867.25$\pm$15.38 \\
2025 02 11 & 60\,717.99 &  11.72 		& $\cdots$ 			& $\cdots$ 			& 536.97$\pm$38.06 	& $\cdots$ 			& $\cdots$ \\
2025 02 13 & 60\,719.45 &  13.17 		& $\cdots$ 			& $\cdots$ 			& $\cdots$ 			& $\cdots$ 			& 772.33$\pm$16.14 \\
2025 02 13 & 60\,719.87 &  13.58 		& $\cdots$ 			& $\cdots$ 			& 435.76$\pm$20.82 	& $\cdots$ 			& $\cdots$ \\
2025 02 14 & 60\,720.85 &  14.56 		& $\cdots$ 			& $\cdots$ 			& 407.69$\pm$39.18 	& $\cdots$ 			& $\cdots$ \\
2025 02 15 & 60\,721.52 &  15.22 		& $\cdots$ 			& $\cdots$ 			& $\cdots$ 			& $\cdots$ 			& 657.50$\pm$12.53 \\
2025 02 16 & 60\,722.03 &  15.73 		& $\cdots$ 			& $\cdots$ 			& 462.41$\pm$23.08 	& $\cdots$ 			& $\cdots$ \\
2025 02 16 & 60\,722.29 &  15.99 		& 145.80$\pm$7.02 	& $\cdots$ 			& $\cdots$ 			& $\cdots$ 			& $\cdots$ \\
2025 02 17 & 60\,723.44 &  17.13 		& $\cdots$ 			& $\cdots$ 			& $\cdots$ 			& $\cdots$ 			& 626.75$\pm$8.88 \\
2025 02 18 & 60\,724.33 &  18.01 		& $\cdots$ 			& $\cdots$ 			& $\cdots$ 			& 576.89$\pm$6.37 	& $\cdots$ \\
2025 02 19 & 60\,725.48 &  19.16 		& $\cdots$ 			& $\cdots$ 			& $\cdots$ 			& $\cdots$ 			& 558.88$\pm$7.04 \\
2025 02 19 & 60\,725.91 &  19.58 		& $\cdots$ 			& $\cdots$ 			& 328.68$\pm$11.38 	& $\cdots$ 			& $\cdots$ \\
2025 02 20 & 60\,726.31 &  19.98 		& $\cdots$ 			& $\cdots$ 			& $\cdots$ 			& 545.06$\pm$5.76 	& $\cdots$ \\
2025 02 20 & 60\,726.35 &  20.02 		& 132.74$\pm$3.50 	& $\cdots$ 			& $\cdots$ 			& $\cdots$ 			& $\cdots$ \\
2025 02 21 & 60\,727.45 &  21.11 		& $\cdots$ 			& $\cdots$ 			& $\cdots$ 			& $\cdots$ 			& 507.75$\pm$9.29 \\
2025 02 21 & 60\,727.92 &  21.58 		& $\cdots$ 			& $\cdots$ 			& 311.28$\pm$10.69 	& $\cdots$ 			& $\cdots$ \\
2025 02 22 & 60\,728.21 &  21.87 		& $\cdots$ 			& $\cdots$ 			& $\cdots$ 			& 471.96$\pm$6.16 	& $\cdots$ \\
2025 02 22 & 60\,728.29 &  21.95 		& 123.52$\pm$4.00 	& $\cdots$ 			& $\cdots$ 			& $\cdots$ 			& $\cdots$ \\
2025 02 22 & 60\,728.51 &  22.16 		& $\cdots$ 			& $\cdots$ 			& $\cdots$ 			& $\cdots$ 			& 497.25$\pm$8.88 \\
2025 02 23 & 60\,729.41 &  23.06 		& $\cdots$ 			& $\cdots$ 			& $\cdots$ 			& $\cdots$ 			& 464.67$\pm$9.83 \\
2025 02 23 & 60\,729.92 &  23.57 		& $\cdots$ 			& $\cdots$ 			& 268.90$\pm$15.48 	& $\cdots$ 			& $\cdots$ \\
2025 02 24 & 60\,730.22 &  23.86 		& $\cdots$ 			& $\cdots$ 			& $\cdots$ 			& 452.80$\pm$5.68 	& $\cdots$ \\
2025 02 24 & 60\,730.31 &  23.95 		& 109.83$\pm$4.82 	& $\cdots$ 			& $\cdots$ 			& $\cdots$ 			& $\cdots$ \\
2025 02 24 & 60\,730.47 &  24.11 		& $\cdots$ 			& $\cdots$ 			& $\cdots$ 			& $\cdots$ 			& 475.25$\pm$8.77 \\
2025 02 25 & 60\,731.42 &  25.05 		& $\cdots$ 			& $\cdots$ 			& $\cdots$ 			& $\cdots$ 			& 447.60$\pm$8.36 \\
2025 02 25 & 60\,731.90 &  25.53 		& $\cdots$ 			& $\cdots$ 			& 277.83$\pm$15.48 	& $\cdots$ 			& $\cdots$ \\
2025 02 26 & 60\,732.20 &  25.83 		& 109.11$\pm$3.59 	& $\cdots$ 			& $\cdots$ 			& $\cdots$ 			& $\cdots$ \\
2025 02 26 & 60\,732.26 &  25.89 		& $\cdots$ 			& $\cdots$ 			& $\cdots$ 			& 427.96$\pm$5.23 & $\cdots$ \\
2025 02 27 & 60\,733.52 &  27.14 		& $\cdots$ 			& $\cdots$ 			& $\cdots$ 			& $\cdots$ 			& 428.50$\pm$8.40 \\
2025 02 27 & 60\,733.89 &  27.51 		& $\cdots$ 			& $\cdots$ 			& 234.53$\pm$10.91 	& $\cdots$ 			& $\cdots$ \\
2025 02 28 & 60\,734.31 &  27.92 		& 101.75$\pm$4.73 	& $\cdots$ 			& $\cdots$ 			& $\cdots$ 			& $\cdots$ \\
2025 02 28 & 60\,734.33 &  27.94 		& $\cdots$ 			& $\cdots$ 			& $\cdots$ 			& 373.79$\pm$8.45 	& $\cdots$ \\
2025 03 01 & 60\,735.41 &  29.02 		& $\cdots$ 			& $\cdots$ 			& $\cdots$ 			& $\cdots$ 			& 398.25$\pm$7.51 \\
2025 03 02 & 60\,736.30 &  29.90 		& $\cdots$ 			& $\cdots$ 			& $\cdots$ 			& $\cdots$ 			& 384.25$\pm$8.82 \\
2025 03 03 & 60\,737.40 &  30.99 		& $\cdots$ 			& $\cdots$ 			& $\cdots$ 			& $\cdots$ 			& 374.50$\pm$7.75 \\
2025 03 05 & 60\,739.24 &  32.82 		& $\cdots$ 			& $\cdots$ 			& $\cdots$ 			& 361.51$\pm$7.11 	& $\cdots$ \\
2025 03 05 & 60\,739.30 &  32.88 		& 100.04$\pm$4.10 	& $\cdots$ 			& $\cdots$ 			& $\cdots$ 			& $\cdots$ \\
2025 03 05 & 60\,739.41 &  32.99 		& $\cdots$ 			& $\cdots$ 			& $\cdots$ 			& $\cdots$ 			& 372.25$\pm$10.62 \\
2025 03 06 & 60\,740.41 &  33.98 		& $\cdots$ 			& $\cdots$ 			& $\cdots$ 			& $\cdots$ 			& 336.25$\pm$10.52 \\
2025 03 07 & 60\,741.40 &  34.97 		& $\cdots$ 			& $\cdots$ 			& $\cdots$ 			& $\cdots$ 			& 319.12$\pm$7.20 \\
2025 03 09 & 60\,743.43 &  36.98 		& $\cdots$ 			& $\cdots$ 			& $\cdots$ 			& $\cdots$ 			& 297.00$\pm$124.46 \\
2025 03 10 & 60\,744.13 &  37.68 		& 80.93$\pm$8.20 	& $\cdots$ 			& $\cdots$ 			& $\cdots$ 			& $\cdots$ \\
2025 03 10 & 60\,744.28 &  37.83 		& $\cdots$ 			& $\cdots$ 			& $\cdots$ 			& 292.31$\pm$6.98 	& $\cdots$ \\
2025 03 10 & 60\,744.99 &  38.53 		& $\cdots$ 			& $\cdots$ 			& 177.62$\pm$21.99 	& $\cdots$ 			& $\cdots$ \\
2025 03 11 & 60\,745.56 &  39.10 		& $\cdots$ 			& $\cdots$ 			& $\cdots$ 			& $\cdots$ 			& 317.00$\pm$16.64 \\
2025 03 11 & 60\,745.89 &  39.42 		& $\cdots$ 			& $\cdots$ 			& 215.75$\pm$36.56 	& $\cdots$ 			& $\cdots$ \\
2025 03 13 & 60\,747.24 &  40.76 		& $\cdots$ 			& $\cdots$ 			& $\cdots$ 			& $\cdots$ 			& 256.00$\pm$82.00 \\
2025 03 14 & 60\,748.39 &  41.91 		& $\cdots$ 			& $\cdots$ 			& $\cdots$ 			& $\cdots$ 			& 293.50$\pm$17.13 \\
2025 03 15 & 60\,749.39 &  42.90 		& $\cdots$ 			& $\cdots$ 			& $\cdots$ 			& $\cdots$ 			& 279.75$\pm$8.49 \\
2025 03 17 & 60\,751.22 &  44.72 		& 91.60$\pm$9.04 	& $\cdots$ 			& $\cdots$ 			& $\cdots$ 			& $\cdots$ \\
2025 03 17 & 60\,751.27 &  44.77 		& $\cdots$ 			& $\cdots$ 			& $\cdots$ 			& 269.25$\pm$5.02 	& $\cdots$ \\
2025 03 17 & 60\,751.39 &  44.89 		& $\cdots$ 			& $\cdots$ 			& $\cdots$ 			& $\cdots$ 			& 252.71$\pm$8.56 \\
2025 03 20 & 60\,754.30 &  47.78 		& 78.91$\pm$4.07 	& $\cdots$ 			& $\cdots$ 			& $\cdots$ 			& $\cdots$ \\
2025 03 20 & 60\,754.36 &  47.84 		& $\cdots$ 			& $\cdots$ 			& $\cdots$ 			& 239.32$\pm$5.05 	& $\cdots$ \\
2025 03 21 & 60\,755.38 &  48.85 		& $\cdots$ 			& 151.67$\pm$21.24 	& $\cdots$ 			& $\cdots$ 			& $\cdots$ \\
2025 03 22 & 60\,756.24 &  49.70 		& 86.61$\pm$3.73 	& $\cdots$ 			& $\cdots$ 			& $\cdots$ 			& $\cdots$ \\
2025 03 22 & 60\,756.30 &  49.76 		& $\cdots$ 			& $\cdots$ 			& $\cdots$ 			& 233.55$\pm$4.59 	& $\cdots$ \\
2025 03 23 & 60\,757.41 &  50.86 		& $\cdots$ 			& $\cdots$ 			& $\cdots$ 			& $\cdots$ 			& 242.00$\pm$8.51 \\
2025 03 24 & 60\,758.26 &  51.71 		& 88.38$\pm$3.42 	& $\cdots$ 			& $\cdots$ 			& $\cdots$ 			& $\cdots$ \\
2025 03 24 & 60\,758.32 &  51.77 		& $\cdots$ 			& $\cdots$ 			& $\cdots$ 			& 231.98$\pm$4.27 	& $\cdots$ \\
2025 03 25 & 60\,759.38 &  52.82 		& $\cdots$ 			& 149.25$\pm$7.64 	& $\cdots$ 			& $\cdots$ 			& $\cdots$ \\
2025 03 26 & 60\,760.25 &  53.68 		& 87.29$\pm$3.02 	& $\cdots$ 			& $\cdots$ 			& $\cdots$ 			& $\cdots$ \\
2025 03 26 & 60\,760.53 &  53.96 		& $\cdots$ 			& $\cdots$ 			& $\cdots$ 			& $\cdots$ 			& 221.00$\pm$12.13 \\
2025 03 27 & 60\,761.36 &  54.79 		& $\cdots$ 			& $\cdots$ 			& $\cdots$ 			& $\cdots$ 			& 227.25$\pm$8.65 \\
\hline\hline
\multicolumn{8}{l}{$^{a}$ Phases are given relative to GOTO $L$-band maximum, MJD = 60\,706.19.} \\
\end{longtable}
}
\end{center}

\begin{center}
\footnotesize{
\begin{longtable}{lcrrrrrr}
\caption{LT photometry of SN~2025qe.}  \\
\hline
\multicolumn{1}{c}{\textbf{Date}}	 &	\multicolumn{1}{c}{\textbf{MJD}} &  \multicolumn{1}{c}{\textbf{Phase}$^{a}$}   & 	\multicolumn{1}{c}{\textbf{$u$}} &	\multicolumn{1}{c}{\textbf{$g$}} &	\multicolumn{1}{c}{\textbf{$r$}} &	\multicolumn{1}{c}{\textbf{$i$}} &	\multicolumn{1}{c}{\textbf{$z$}}  \\ 
\multicolumn{1}{l}{}	             &	\multicolumn{1}{c}{}     &  \multicolumn{1}{c}{\textbf{(days)}}	               & \multicolumn{1}{c}{\textbf{($\mu$Jy)}}	   &	\multicolumn{1}{c}{\textbf{($\mu$Jy)}} &  \multicolumn{1}{c}{\textbf{($\mu$Jy)}} &  \multicolumn{1}{c}{\textbf{($\mu$Jy)}} &  \multicolumn{1}{c}{\textbf{($\mu$Jy)}}  \\ \hline\hline
\label{tab:25qe_lt_phot}
\endfirsthead
\multicolumn{8}{c}%
{{\tablename\ \thetable{} -- continued from previous page}} \\
\hline
\multicolumn{1}{c}{\textbf{Date}}	 &	\multicolumn{1}{c}{\textbf{MJD}} &  \multicolumn{1}{c}{\textbf{Phase}$^{a}$}   & 	\multicolumn{1}{c}{\textbf{$u$}} &	\multicolumn{1}{c}{\textbf{$g$}} &	\multicolumn{1}{c}{\textbf{$r$}} &	\multicolumn{1}{c}{\textbf{$i$}} &	\multicolumn{1}{c}{\textbf{$z$}} \\ 
\multicolumn{1}{l}{}	             &	\multicolumn{1}{c}{}     &  \multicolumn{1}{c}{\textbf{(days)}}	               & \multicolumn{1}{c}{\textbf{($\mu$Jy)}}	   &	\multicolumn{1}{c}{\textbf{($\mu$Jy)}} &  \multicolumn{1}{c}{\textbf{($\mu$Jy)}} &  \multicolumn{1}{c}{\textbf{($\mu$Jy)}} &  \multicolumn{1}{c}{\textbf{($\mu$Jy)}}  \\ \hline\hline
\endhead

\hline \multicolumn{8}{|r|}{{Continued on next page}} \\ \hline
\endfoot

\endlastfoot

2025 01 29 & 60\,704.98 & $-$1.20 & 437.32$\pm$5.64 & 1033.71$\pm$6.66 & 1077.46$\pm$5.95 & 884.71$\pm$3.26 & 778.39$\pm$12.90 \\
2025 01 30 & 60\,705.98 & $-$0.21 & 384.06$\pm$6.72 & 1019.53$\pm$7.51 & 1098.50$\pm$7.08 & 902.82$\pm$3.33 & 796.53$\pm$3.67 \\
2025 01 31 & 60\,706.97 & 0.77 & 327.79$\pm$6.04 & 975.44$\pm$7.19 & 1109.69$\pm$6.13 & 941.02$\pm$3.47 & 827.18$\pm$4.57 \\
2025 02 02 & 60\,708.96 & 2.75 & 223.87$\pm$5.77 & 865.37$\pm$7.17 & 1130.32$\pm$6.25 & 982.65$\pm$3.62 & 887.97$\pm$4.09 \\
2025 02 03 & 60\,709.95 & 3.74 & 180.14$\pm$4.98 & 815.83$\pm$7.51 & 1101.54$\pm$6.09 & 1001.84$\pm$3.69 & 912.01$\pm$4.20 \\
2025 02 04 & 60\,710.96 & 4.73 & 142.30$\pm$4.98 & 697.59$\pm$6.43 & 1108.66$\pm$6.13 & 1024.24$\pm$3.77 & 928.11$\pm$4.27 \\
2025 02 05 & 60\,711.99 & 5.76 & 115.13$\pm$6.79 & 627.48$\pm$8.09 & 1081.43$\pm$6.97 & 1031.81$\pm$2.85 & 939.29$\pm$3.46 \\
2025 02 06 & 60\,712.95 & 6.72 & 90.03$\pm$6.14 & 556.67$\pm$7.69 & 1054.87$\pm$6.80 & 1035.62$\pm$3.82 & 920.45$\pm$4.24 \\
2025 02 08 & 60\,714.94 & 8.69 & 41.11$\pm$9.66 & 440.96$\pm$10.97 & 955.87$\pm$9.68 & 1015.78$\pm$4.68 & 894.54$\pm$9.89 \\
2025 02 10 & 60\,716.94 & 10.67 & $\cdots$ & 325.39$\pm$8.39 & 847.23$\pm$7.80 & 947.11$\pm$5.23 & 855.85$\pm$7.88 \\
2025 02 12 & 60\,718.96 & 12.68 & 41.80$\pm$11.82 & 238.56$\pm$14.94 & 748.17$\pm$8.96 & 843.33$\pm$28.74 & 792.87$\pm$37.97 \\
2025 02 13 & 60\,719.98 & 13.70 & 31.83$\pm$8.03 & 234.86$\pm$9.52 & 700.81$\pm$7.10 & 809.84$\pm$35.06 & 755.09$\pm$46.60 \\
2025 02 15 & 60\,721.94 & 15.64 & 21.14$\pm$4.97 & 200.08$\pm$5.71 & 628.64$\pm$5.21 & 734.51$\pm$12.85 & 682.97$\pm$39.63 \\
2025 02 16 & 60\,722.93 & 16.63 & 22.06$\pm$3.96 & 180.14$\pm$5.14 & 589.39$\pm$4.89 & 701.46$\pm$3.88 & 657.66$\pm$50.88 \\
\hline\hline
\multicolumn{8}{l}{$^{a}$ Phases are given relative to GOTO $L$-band maximum, MJD = 60\,706.19.} \\
\end{longtable}
}
\end{center}

\begin{center}
\footnotesize{
\begin{longtable}{lcrrrrr}
\caption{pt5m photometry of SN~2025qe.}  \\
\hline
\multicolumn{1}{c}{\textbf{Date}}	 &	\multicolumn{1}{c}{\textbf{MJD}} &  \multicolumn{1}{c}{\textbf{Phase}$^{a}$}   & 	\multicolumn{1}{c}{\textbf{$B$}} &	\multicolumn{1}{c}{\textbf{$V$}} &	\multicolumn{1}{c}{\textbf{$R$}} &	\multicolumn{1}{c}{\textbf{$I$}}  \\ 
\multicolumn{1}{l}{}	             &	\multicolumn{1}{c}{}     &  \multicolumn{1}{c}{\textbf{(days)}}	               & \multicolumn{1}{c}{\textbf{($\mu$Jy)}}	   &	\multicolumn{1}{c}{\textbf{($\mu$Jy)}} &  \multicolumn{1}{c}{\textbf{($\mu$Jy)}} &  \multicolumn{1}{c}{\textbf{($\mu$Jy)}}  \\ \hline\hline
\label{tab:25qe_pt5m_phot}
\endfirsthead
\multicolumn{7}{c}%
{{\tablename\ \thetable{} -- continued from previous page}} \\
\hline
\multicolumn{1}{c}{\textbf{Date}}	 &	\multicolumn{1}{c}{\textbf{MJD}} &  \multicolumn{1}{c}{\textbf{Phase}$^{a}$}   & 	\multicolumn{1}{c}{\textbf{$B$}} &	\multicolumn{1}{c}{\textbf{$V$}} &	\multicolumn{1}{c}{\textbf{$R$}} &	\multicolumn{1}{c}{\textbf{$I$}}  \\ 
\multicolumn{1}{l}{}	             &	\multicolumn{1}{c}{}     &  \multicolumn{1}{c}{\textbf{(days)}}	               & \multicolumn{1}{c}{\textbf{($\mu$Jy)}}	   &	\multicolumn{1}{c}{\textbf{($\mu$Jy)}} &  \multicolumn{1}{c}{\textbf{($\mu$Jy)}} &  \multicolumn{1}{c}{\textbf{($\mu$Jy)}}  \\ \hline\hline
\endhead

\hline \multicolumn{7}{|r|}{{Continued on next page}} \\ \hline
\endfoot

\endlastfoot

2025 01 25 & 60\,700.88 & $-$5.27 & $\cdots$ & 870.20$\pm$125.00 & $\cdots$ & $\cdots$ \\
2025 01 26 & 60\,701.99 & $-$4.17 & 940.20$\pm$70.10 & $\cdots$ & $\cdots$ & $\cdots$ \\
2025 01 27 & 60\,702.00 & $-$4.16 & $\cdots$ & 967.40$\pm$64.20 & $\cdots$ & $\cdots$ \\
2025 01 27 & 60\,702.02 & $-$4.15 & $\cdots$ & $\cdots$ & 929.90$\pm$73.70 & $\cdots$ \\
2025 01 27 & 60\,702.92 & $-$3.25 & 899.60$\pm$116.80 & $\cdots$ & $\cdots$ & $\cdots$ \\
2025 01 27 & 60\,702.93 & $-$3.23 & $\cdots$ & 942.80$\pm$63.40 & $\cdots$ & $\cdots$ \\
2025 01 27 & 60\,702.94 & $-$3.22 & $\cdots$ & $\cdots$ & 933.30$\pm$65.30 & $\cdots$ \\
2025 01 28 & 60\,703.93 & $-$2.25 & 968.30$\pm$243.50 & $\cdots$ & $\cdots$ & $\cdots$ \\
2025 01 28 & 60\,703.94 & $-$2.23 & $\cdots$ & 1007.50$\pm$92.80 & $\cdots$ & $\cdots$ \\
2025 01 28 & 60\,703.96 & $-$2.22 & 850.40$\pm$117.50 & 1006.50$\pm$84.40 & 1031.90$\pm$57.00 & 1181.50$\pm$59.90 \\
2025 01 29 & 60\,704.92 & $-$1.26 & 772.70$\pm$70.50 & $\cdots$ & $\cdots$ & $\cdots$ \\
2025 02 02 & 60\,708.89 & 2.68 & 541.00$\pm$66.80 & $\cdots$ & $\cdots$ & $\cdots$ \\
2025 02 03 & 60\,709.91 & 3.69 & 539.00$\pm$50.10 & 1086.50$\pm$196.10 & 1231.50$\pm$90.70 & 1364.00$\pm$93.00 \\
2025 02 05 & 60\,711.97 & 5.74 & 358.10$\pm$52.40 & 767.80$\pm$66.50 & $\cdots$ & $\cdots$ \\
\hline\hline
\multicolumn{7}{l}{$^{a}$ Phases are given relative to GOTO $L$-band maximum, MJD = 60\,706.19.} \\
\end{longtable}
}
\end{center}

\begin{center}
\footnotesize{
\begin{longtable}{lcrrrr}
\caption{TTT photometry of SN~2025qe.}  \\
\hline
\multicolumn{1}{c}{\textbf{Date}}	 &	\multicolumn{1}{c}{\textbf{MJD}} &  \multicolumn{1}{c}{\textbf{Phase}$^{a}$}   & 	\multicolumn{1}{c}{\textbf{$g$}} &	\multicolumn{1}{c}{\textbf{$r$}} &	\multicolumn{1}{c}{\textbf{$i$}}    \\ 
\multicolumn{1}{l}{}	             &	\multicolumn{1}{c}{}     &  \multicolumn{1}{c}{\textbf{(days)}}	               & \multicolumn{1}{c}{\textbf{($\mu$Jy)}}	   &	\multicolumn{1}{c}{\textbf{($\mu$Jy)}} &  \multicolumn{1}{c}{\textbf{($\mu$Jy)}}   \\ \hline\hline
\label{tab:25qe_ttt_phot}
\endfirsthead
\multicolumn{6}{c}%
{{\tablename\ \thetable{} -- continued from previous page}} \\
\hline
\multicolumn{1}{c}{\textbf{Date}}	 &	\multicolumn{1}{c}{\textbf{MJD}} &  \multicolumn{1}{c}{\textbf{Phase}$^{a}$}   & 	\multicolumn{1}{c}{\textbf{$g$}} &	\multicolumn{1}{c}{\textbf{$r$}} &	\multicolumn{1}{c}{\textbf{$i$}}    \\ 
\multicolumn{1}{l}{}	             &	\multicolumn{1}{c}{}     &  \multicolumn{1}{c}{\textbf{(days)}}	               & \multicolumn{1}{c}{\textbf{($\mu$Jy)}}	   &	\multicolumn{1}{c}{\textbf{($\mu$Jy)}} &  \multicolumn{1}{c}{\textbf{($\mu$Jy)}}   \\ \hline\hline
\endhead

\hline \multicolumn{6}{|r|}{{Continued on next page}} \\ \hline
\endfoot

\endlastfoot
2025 01 24 & 60\,699.83 & $-$6.32 & 833.30$\pm$26.86 & 813.58$\pm$35.97 & 631.54$\pm$47.12 \\
2025 01 25 & 60\,700.83 & $-$5.32 & 917.06$\pm$25.34 & 882.27$\pm$34.13 & $\cdots$ \\
2025 01 26 & 60\,701.83 & $-$4.33 & 950.60$\pm$27.14 & 999.08$\pm$38.65 & 741.31$\pm$51.21 \\
2025 01 28 & 60\,703.83 & $-$2.34 & 996.32$\pm$25.69 & 1075.47$\pm$36.65 & 869.36$\pm$48.84 \\
2025 01 29 & 60\,704.83 & $-$1.35 & 1041.36$\pm$28.77 & 1085.43$\pm$38.99 & 935.84$\pm$56.89 \\
2025 02 01 & 60\,707.15 & 0.95 & 930.68$\pm$28.29 & 1111.73$\pm$38.91 & 930.68$\pm$55.72 \\
2025 02 01 & 60\,707.90 & 1.70 & 878.21$\pm$28.31 & 1117.89$\pm$38.10 & 923.85$\pm$51.90 \\
2025 02 05 & 60\,711.02 & 4.80 & 633.87$\pm$21.02 & 1103.57$\pm$34.56 & $\cdots$ \\
\hline\hline
\multicolumn{6}{l}{$^{a}$ Phases are given relative to GOTO $L$-band maximum, MJD = 60\,706.19.} \\
\end{longtable}
}
\end{center}

\clearpage

\section{Spectroscopic logs}
\label{appendix_spec}

\begin{table*}
\begin{center}
\caption{Spectroscopy of SN~2024bfu.}
\label{tab:spec_24bfu}
\begin{tabular}{lcrcccc}
\hline
\textbf{Date} & \textbf{MJD} & \textbf{Phase$^{a}$} & \textbf{Instrument} & \textbf{Grism} & \textbf{Wavelength coverage (\AA)} & \textbf{Resolution (\AA)} \\
\hline
\hline
2024 Jan 31 & 60\,341.30 & $-$6.25      & NTT+EFOSC2 & Gr11       & 3\,400 -- \phn{}7\,500     & 4.08 \\
2024 Feb 04 & 60\,344.84 & $-$2.83      & INT+IDS    & R150V      & 3\,800 -- 10\,000          & 4.11 \\
2024 Feb 13 & 60\,353.04 & 5.09         & INT+IDS    & R150V      & 3\,800 -- \phn{}8\,400     & 4.10 \\
2024 Mar 05 & 60\,375.13 & 26.41        & NTT+EFOSC2 & Gr11+Gr16  & 3\,400 -- 10\,000          & 4.21 \\
2024 Mar 13 & 60\,383.17 & 34.17        & NTT+EFOSC2 & Gr11+Gr16  & 3\,400 -- 10\,000          & 4.21 \\
2024 Mar 31 & 60\,401.04 & 51.42        & NTT+EFOSC2 & Gr13       & 3\,700 -- \phn{}9\,300     & 5.52 \\
\hline
\hline
\multicolumn{6}{l}{$^{a}$ Phases are given relative to GOTO $L$-band maximum, MJD = 60\,347.77.} \\
\end{tabular}
\end{center}
\end{table*}

\begin{table*}
\begin{center}
\caption{Spectroscopy of SN~2025qe.}
\label{tab:spec_25qe}
\begin{tabular}{lcrccc}
\hline
\textbf{Date} & \textbf{MJD} & \textbf{Phase$^{a}$} & \textbf{Instrument} & \textbf{Wavelength coverage (\AA)} & \textbf{Resolution (\AA)} \\
\hline
\hline
2025 Jan 18 & 60\,693.80 & $-$12.30 & Lijiang-2.4m+YFOSC $^{b}$   &        3\,600 -- 8\,900 & 2.86  \\
2025 Jan 20 & 60\,695.01 & $-$11.10 & LT+SPRAT              &        4\,000 -- 8\,100     & 4.65 \\
2025 Jan 26 & 60\,701.03 & $-$5.12  & LT+SPRAT               &        4\,000 -- 8\,100     & 4.65 \\
2025 Jan 28 & 60\,703.98 & $-$2.19  & LT+SPRAT               &        4\,000 -- 8\,100     & 4.65 \\
2025 Jan 30 & 60\,705.99 & $-$0.20  & LT+SPRAT               &        4\,000 -- 8\,100     & 4.65 \\
2025 Feb 02 & 60\,708.97 & 2.76     & LT+SPRAT                  &        4\,000 -- 8\,100     & 4.65 \\
2025 Feb 04 & 60\,710.97 & 4.75     & LT+SPRAT                  &        4\,000 -- 8\,100     & 4.65 \\
2025 Feb 06 & 60\,712.97 & 6.73     & LT+SPRAT                  &        4\,000 -- 8\,100     & 4.65 \\
2025 Feb 08 & 60\,714.95 & 8.70     & LT+SPRAT                  &        4\,000 -- 8\,100    & 4.65 \\
2025 Feb 10 & 60\,716.94 & 10.68    & LT+SPRAT                  &       4\,000 -- 8\,100    & 4.65 \\
2025 Feb 12 & 60\,718.97 & 12.69    & LT+SPRAT                  &       4\,000 -- 8\,100    & 4.65 \\
2025 Feb 15 & 60\,721.95 & 15.65    & LT+SPRAT                  &       4\,000 -- 8\,100    & 4.65 \\
2025 Feb 20 & 60\,726.95 & 20.62    & LT+SPRAT                  &       4\,000 -- 8\,100    & 4.65 \\
2025 Feb 28 & 60\,734.90 & 28.51    & LT+SPRAT                  &       4\,000 -- 8\,100    & 4.65 \\
2025 Mar 28 & 60\,762.99 & 56.41    & CAHA2.2+CAFOS             &       3\,800 -- 8\,700    & 4.50 \\
\hline
\hline
\multicolumn{5}{l}{$^{a}$ Phases are given relative to GOTO $L$-band maximum, MJD = 60\,706.19.} \\
\multicolumn{5}{l}{$^{b}$ Obtained from the TNS.} \\
\end{tabular}
\end{center}
\end{table*}

\section{Sample properties and plots}
\label{appendix_props}

\begin{landscape}
\begin{table}
\centering
\caption{SNe~Iax sample properties.}
\label{tab:sample_props}
\resizebox{!}{4.3cm}{
\begin{tabular}{lllllllllllllll}
\hline
\textbf{Name} & \textbf{$z$} & \textbf{First light} & \multicolumn{4}{c}{\textbf{Peak absolute magnitudes}} &  \multicolumn{4}{c}{\textbf{Rise times}} &  \multicolumn{4}{c}{\textbf{Rise indices}}   \\
              &       & \textbf{$t_0$}       & \textbf{$M_g$} & \textbf{$M_L$} & \textbf{$M_r$}  & \textbf{$M_o$} & \textbf{$t_g$}  & \textbf{$t_L$} & \textbf{$t_r$}   & \textbf{$t_o$} & \textbf{$\alpha_g$}  & \textbf{$\alpha_L$} & \textbf{$\alpha_r$}   & \textbf{$\alpha_o$}   \\

\hline
\hline
SN2018cxk & 0.030 & 58\,288.65$_{-0.34}^{+0.18}$ & $-$17.06$\pm$0.15 & $\cdots$ & $-$16.99$\pm$0.15 & $\cdots$ & 11.06$_{-0.18}^{+0.34}$ & $\cdots$ & 14.67$_{-0.22}^{+0.36}$ & $\cdots$ & 1.02$_{-0.18}^{+0.11}$ & $\cdots$ & 1.03$_{-0.15}^{+0.10}$ & $\cdots$ \\ [5pt] 
SN2020sck & 0.017 & 59\,085.52$_{-0.18}^{+0.20}$ & $-$18.16$\pm$0.15 & $\cdots$ & $-$17.99$\pm$0.15 & $\cdots$ & 13.90$_{-0.20}^{+0.18}$ & $\cdots$ & 16.99$_{-0.34}^{+0.33}$ & $\cdots$ & 1.33$_{-0.11}^{+0.11}$ & $\cdots$ & 1.43$_{-0.18}^{+0.20}$ & $\cdots$ \\ [5pt] 
SN2020udy & 0.017 & 59\,115.42$_{-0.23}^{+0.26}$ & $-$18.08$\pm$0.15 & $\cdots$ & $-$18.43$\pm$0.15 & $\cdots$ & 15.45$_{-0.26}^{+0.23}$ & $\cdots$ & 20.80$_{-0.53}^{+0.52}$ & $\cdots$ & 1.19$_{-0.13}^{+0.13}$ & $\cdots$ & 1.17$_{-0.08}^{+0.09}$ & $\cdots$ \\ [5pt] 
SN2021jun & 0.040 & 59\,312.85$_{-1.17}^{+0.76}$ & $-$16.96$\pm$0.16 & $\cdots$ & $-$17.72$\pm$0.15 & $\cdots$ & 14.17$_{-0.75}^{+1.14}$ & $\cdots$ & 19.85$_{-0.76}^{+1.14}$ & $\cdots$ & 1.38$_{-0.38}^{+0.35}$ & $\cdots$ & 1.46$_{-0.37}^{+0.27}$ & $\cdots$ \\ [5pt] 
SN2021mry & 0.027 & 59\,350.57$_{-0.54}^{+0.60}$ & $-$17.84$\pm$0.15 & $\cdots$ & $-$18.11$\pm$0.15 & $\cdots$ & 13.41$_{-0.59}^{+0.53}$ & $\cdots$ & 17.93$_{-0.60}^{+0.54}$ & $\cdots$ & 1.18$_{-0.19}^{+0.20}$ & $\cdots$ & 1.35$_{-0.23}^{+0.25}$ & $\cdots$ \\ [5pt] 
SN2022yog & 0.023 & 59\,870.71$_{-1.28}^{+1.22}$ & $-$16.22$\pm$0.15 & $\cdots$ & $-$17.15$\pm$0.15 & $-$17.11$\pm$0.17 & 12.51$_{-1.28}^{+1.34}$ & $\cdots$ & 17.42$_{-1.21}^{+1.27}$ & 18.52$_{-1.30}^{+1.35}$ & 1.12$_{-0.52}^{+0.52}$ & $\cdots$ & 1.43$_{-0.42}^{+0.48}$ & 0.72$_{-0.29}^{+0.26}$ \\ [5pt] 
SN2023umr & 0.027 & 60\,222.55$_{-0.77}^{+0.81}$ & $\cdots$ & $\cdots$ & $\cdots$ & $-$18.01$\pm$0.16 & $\cdots$ & $\cdots$ & $\cdots$ & 20.43$_{-0.95}^{+0.93}$ & $\cdots$ & $\cdots$ & $\cdots$ & 1.18$_{-0.35}^{+0.29}$ \\ [5pt] 
SN2023vez & 0.026 & 60\,229.43$_{-0.74}^{+0.68}$ & $-$18.00$\pm$0.15 & $\cdots$ & $\cdots$ & $-$18.13$\pm$0.16 & 14.17$_{-0.67}^{+0.73}$ & $\cdots$ & $\cdots$ & 21.11$_{-0.73}^{+0.78}$ & 1.50$_{-0.29}^{+0.31}$ & $\cdots$ & $\cdots$ & 1.44$_{-0.34}^{+0.35}$ \\ [5pt] 
SN2023vjq & 0.051 & 60\,218.85$_{-2.45}^{+1.77}$ & $-$18.12$\pm$0.15 & $\cdots$ & $\cdots$ & $\cdots$ & 16.92$_{-1.73}^{+2.37}$ & $\cdots$ & $\cdots$ & $\cdots$ & 1.72$_{-0.42}^{+0.48}$ & $\cdots$ & $\cdots$ & $\cdots$ \\ [5pt] 
SN2023woh & 0.047 & 60\,241.06$_{-4.54}^{+3.13}$ & $-$17.95$\pm$0.15 & $\cdots$ & $-$18.17$\pm$0.17 & $-$18.24$\pm$0.16 & 17.16$_{-3.00}^{+4.34}$ & $\cdots$ & 20.16$_{-3.19}^{+4.48}$ & 22.16$_{-3.07}^{+4.39}$ & 1.56$_{-0.70}^{+0.69}$ & $\cdots$ & 1.66$_{-0.65}^{+0.70}$ & 1.88$_{-0.59}^{+0.71}$ \\ [5pt] 
SN2024bfu & 0.036 & 60\,332.87$_{-1.72}^{+0.67}$ & $\cdots$ & $-$18.05$\pm$0.15 & $\cdots$ & $-$18.35$\pm$0.15 & $\cdots$ & 14.38$_{-0.69}^{+1.68}$ & $\cdots$ & 19.10$_{-0.75}^{+1.71}$ & $\cdots$ & 0.86$_{-0.65}^{+0.41}$ & $\cdots$ & 1.23$_{-0.52}^{+0.53}$ \\ [5pt] 
SN2024pxl & 0.006 & 60\,512.86$_{-0.16}^{+0.28}$ & $\cdots$ & $\cdots$ & $\cdots$ & $-$16.81$\pm$0.15 & $\cdots$ & $\cdots$ & $\cdots$ & 17.02$_{-0.28}^{+0.17}$ & $\cdots$ & $\cdots$ & $\cdots$ & 0.93$_{-0.06}^{+0.08}$ \\ [5pt] 
SN2024vjm & 0.003 & 60\,565.63$_{-0.45}^{+0.39}$ & $\cdots$ & $-$13.19$\pm$0.15 & $\cdots$ & $\cdots$ & $\cdots$ & 9.66$_{-0.40}^{+0.47}$ & $\cdots$ & $\cdots$ & $\cdots$ & 0.94$_{-0.34}^{+0.32}$ & $\cdots$ & $\cdots$ \\ [5pt] 
SN2025ay  & 0.020 & 60\,676.95$_{-1.12}^{+0.98}$ & $\cdots$ & $\cdots$ & $\cdots$ & $-$18.03$\pm$0.15 & $\cdots$ & $\cdots$ & $\cdots$ & 26.22$_{-2.17}^{+2.23}$ & $\cdots$ & $\cdots$ & $\cdots$ & 1.38$_{-0.50}^{+0.53}$ \\ [5pt] 
\hline
\end{tabular}
}
\end{table}
\end{landscape}

\twocolumn
\begin{figure}
\centering
\includegraphics[width=\columnwidth]{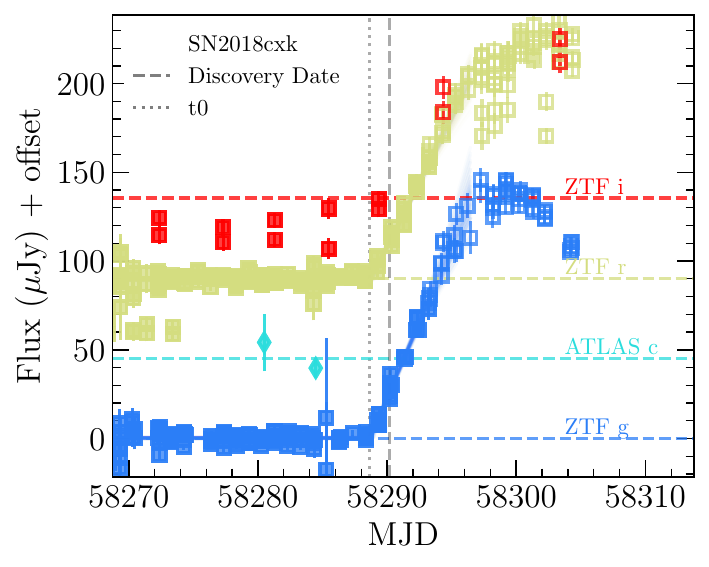}
\caption{As in Fig.~\ref{fig:lc} for SN~2018cxk.}
\label{fig:fit_18cxk}
\centering
\end{figure}

\begin{figure}
\centering
\includegraphics[width=\columnwidth]{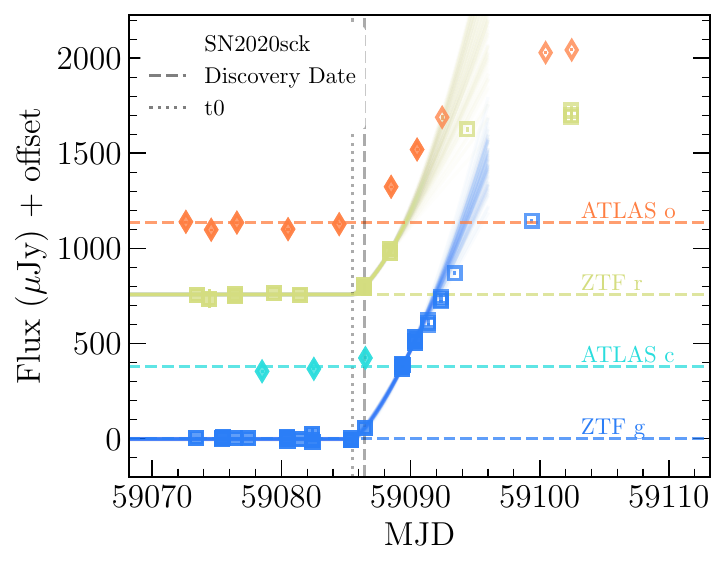}
\caption{As in Fig.~\ref{fig:lc} for SN~2020sck.}
\label{fig:fit_20sck}
\centering
\end{figure}

\begin{figure}
\centering
\includegraphics[width=\columnwidth]{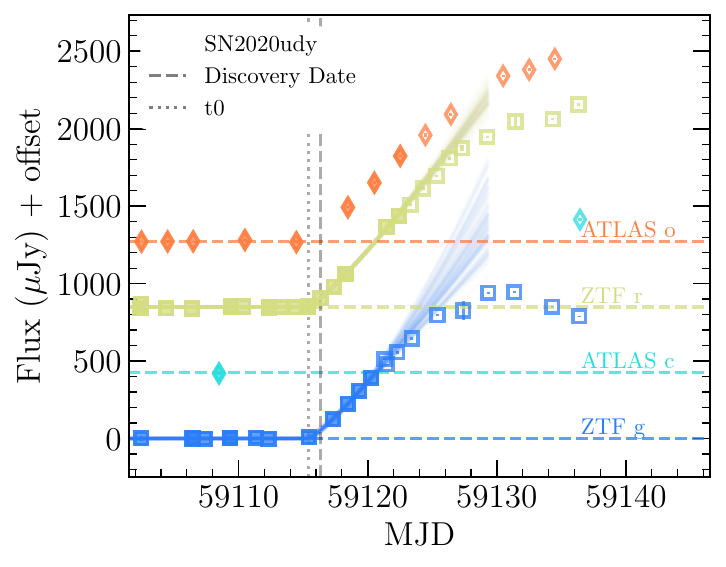}
\caption{As in Fig.~\ref{fig:lc} for SN~2020udy.}
\label{fig:fit_20udy}
\centering
\end{figure}

\begin{figure}
\centering
\includegraphics[width=\columnwidth]{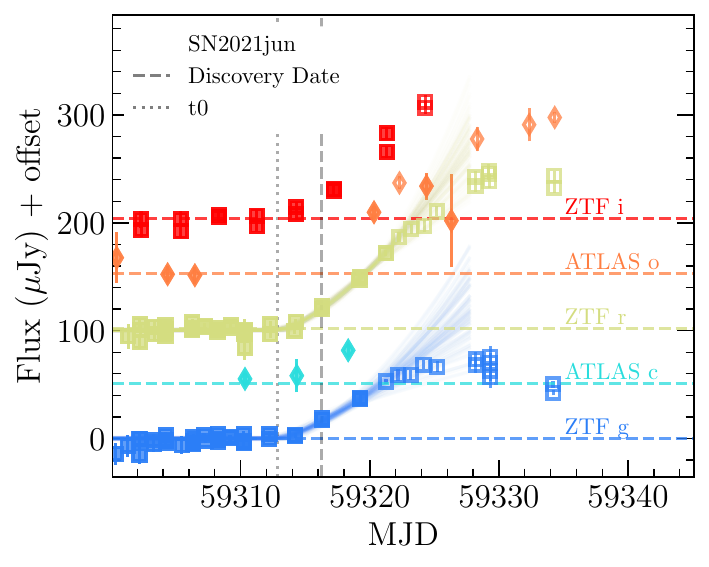}
\caption{As in Fig.~\ref{fig:lc} for SN~2021jun.}
\label{fig:fit_21jun}
\centering
\end{figure}

\begin{figure}
\centering
\includegraphics[width=\columnwidth]{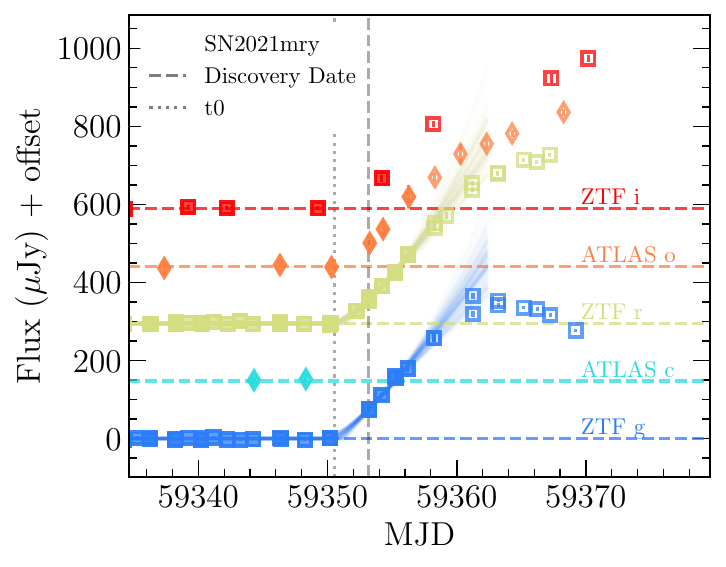}
\caption{As in Fig.~\ref{fig:lc} for SN~2021mry.}
\label{fig:fit_21mry}
\centering
\end{figure}

\begin{figure}
\centering
\includegraphics[width=\columnwidth]{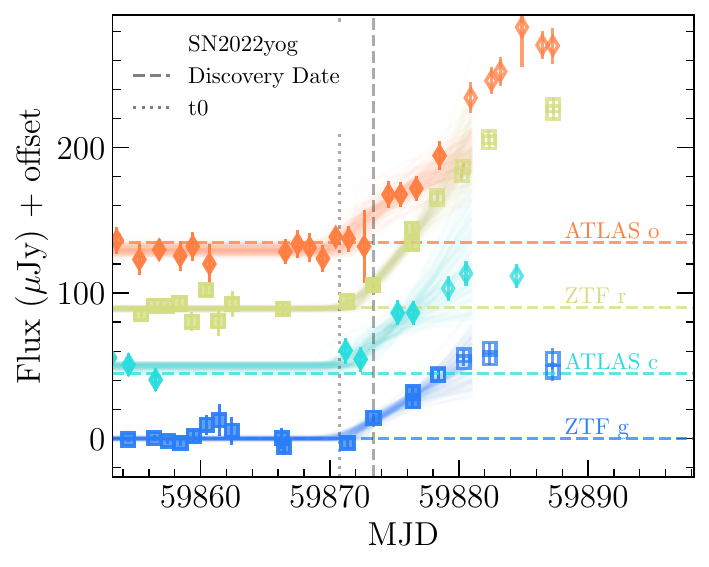}
\caption{As in Fig.~\ref{fig:lc} for SN~2022yog.}
\label{fig:fit_22yog}
\centering
\end{figure}

\begin{figure}
\centering
\includegraphics[width=\columnwidth]{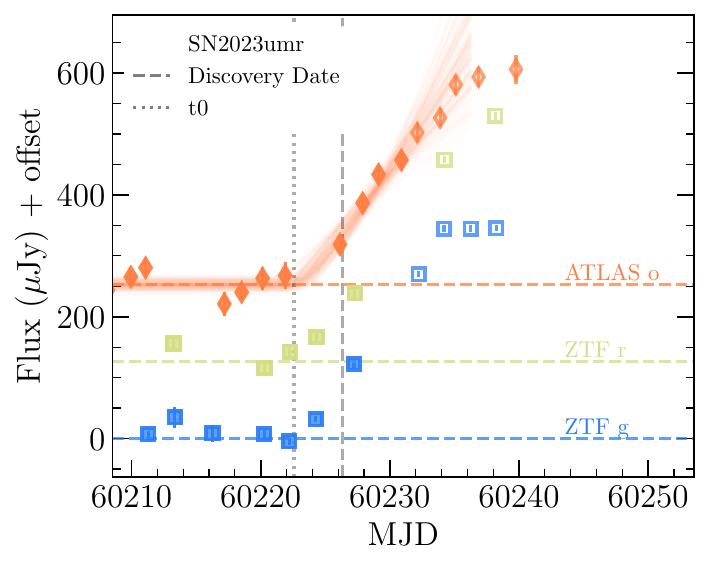}
\caption{As in Fig.~\ref{fig:lc} for SN~2023umr.}
\label{fig:fit_23umr}
\centering
\end{figure}

\begin{figure}
\centering
\includegraphics[width=\columnwidth]{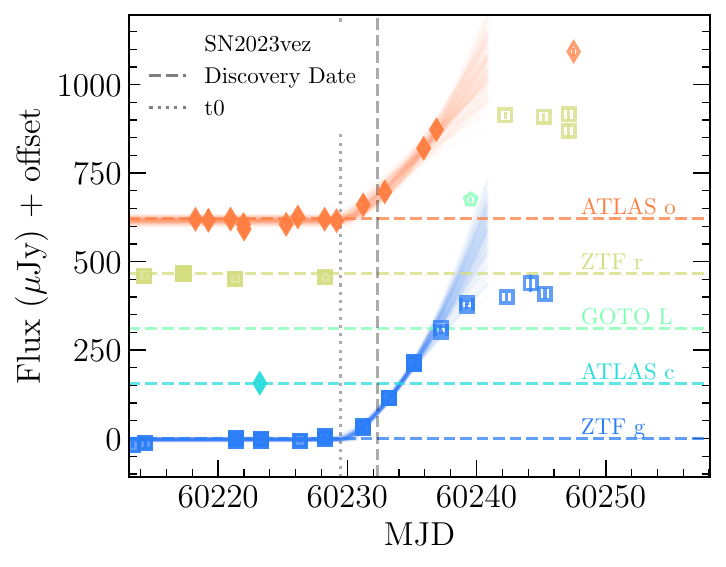}
\caption{As in Fig.~\ref{fig:lc} for SN~2023vez.}
\label{fig:fit_23vez}
\centering
\end{figure}

\begin{figure}
\centering
\includegraphics[width=\columnwidth]{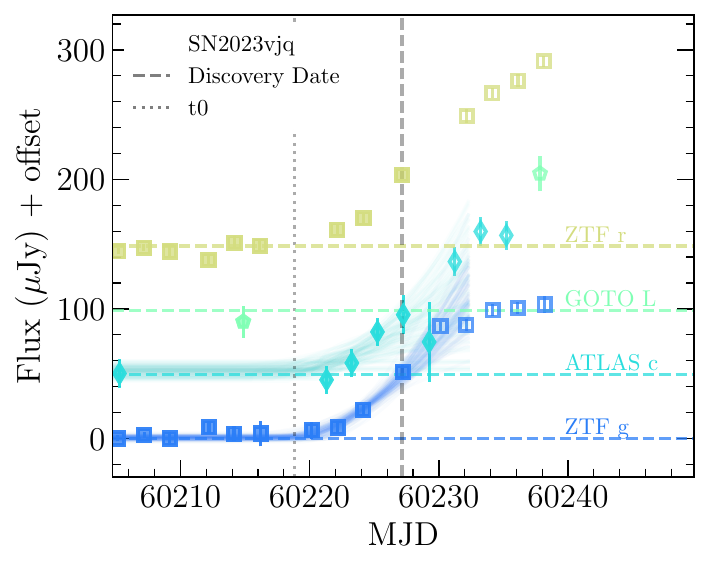}
\caption{As in Fig.~\ref{fig:lc} for SN~2023vjq.}
\label{fig:fit_23vjq}
\centering
\end{figure}

\begin{figure}
\centering
\includegraphics[width=\columnwidth]{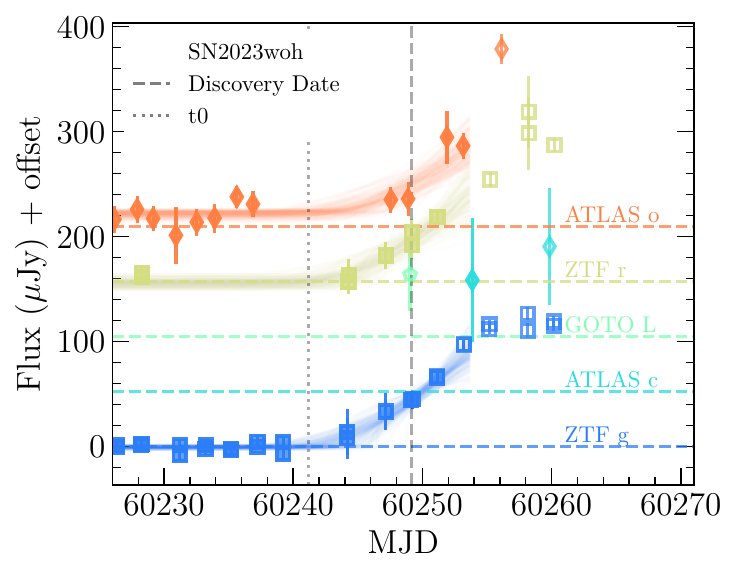}
\caption{As in Fig.~\ref{fig:lc} for SN~2023woh.}
\label{fig:fit_23woh}
\centering
\end{figure}

\begin{figure}
\centering
\includegraphics[width=\columnwidth]{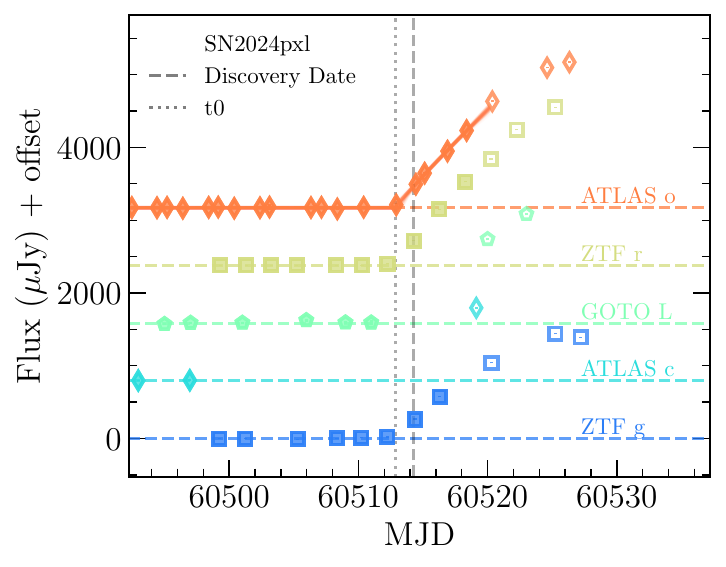}
\caption{As in Fig.~\ref{fig:lc} for SN~2024pxl.}
\label{fig:fit_24pxl}
\centering
\end{figure}

\begin{figure}
\centering
\includegraphics[width=\columnwidth]{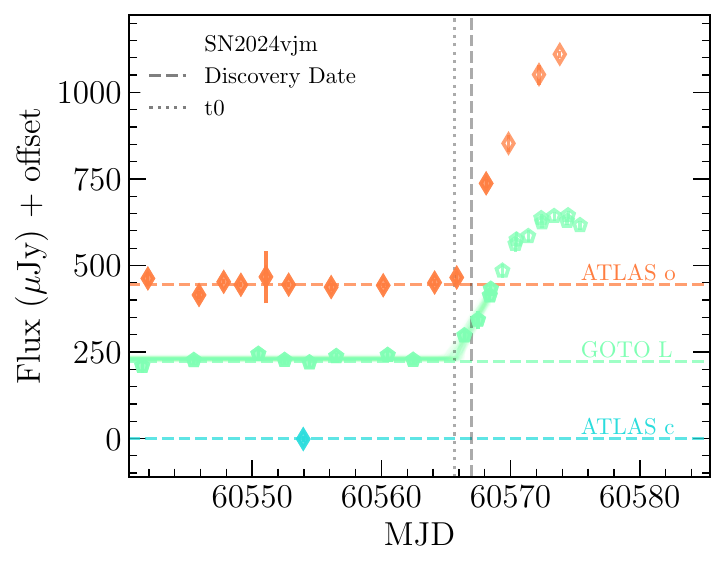}
\caption{As in Fig.~\ref{fig:lc} for SN~2024vjm.}
\label{fig:fit_24vjm}
\centering
\end{figure}

\begin{figure}
\centering
\includegraphics[width=\columnwidth]{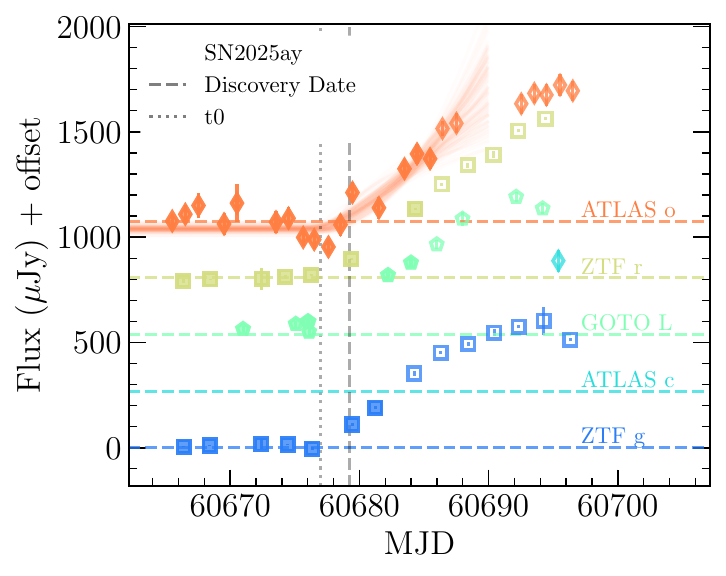}
\caption{As in Fig.~\ref{fig:lc} for SN~2025ay.}
\label{fig:fit_25ay}
\centering
\end{figure}

\onecolumn

\newpage
\footnotesize
\it
\noindent
$^{1}$Department of Physics, University of Warwick, Gibbet Hill Road, Coventry CV4 7AL, UK \\ 
$^{2}$Astrophysics Research Cluster, School of Mathematical and Physical Sciences, University of Sheffield, Sheffield S3 7RH, UK \\
$^{3}$Institut d'Estudis Espacials de Catalunya (IEEC), 08860 Castelldefels (Barcelona), Spain \\
$^{4}$Institute of Space Sciences (ICE-CSIC), Campus UAB, Carrer de Can Magrans, s/n, E-08193 Barcelona, Spain \\
$^{5}$European Southern Observatory, Alonso de C\'ordova 3107, Casilla 19, Santiago, Chile \\
$^{6}$Millennium Institute of Astrophysics MAS, Nuncio Monsenor Sotero Sanz 100, Off. 104, Providencia, Santiago, Chile \\
$^{7}$Centre for Advanced Instrumentation, University of Durham, DH1 3LE Durham, UK \\
$^{8}$Graduate Institute of Astronomy, National Central University, 300 Jhongda Road, 32001 Jhongli, Taiwan \\
$^{9}$Instituto de Astrof\'{\i}sica de Canarias, E-38205 La Laguna, Tenerife, Spain \\
$^{10}$Instituto de Astrof\'isica e Ci\^encias do Espaço, Faculdade de Ci\^encias, Universidade de Lisboa, Ed. C8, Campo Grande, 1749-016 Lisbon, Portugal \\
$^{11}$Astronomical Observatory, University of Warsaw, Al. Ujazdowskie 4, 00-478 Warszawa, Poland \\
$^{12}$Cardiff Hub for Astrophysics Research and Technology, School of Physics \& Astronomy, Cardiff University, Queens Buildings, The Parade, Cardiff, CF24 3AA, UK \\
$^{13}$Institute of Astronomy and Kavli Institute for Cosmology, University of Cambridge, Madingley Road, Cambridge CB3 0HA, UK \\
$^{14}$Department of Physics, Royal Holloway, University of London, Egham, TW20 0EX, UK \\
$^{15}$DTU Space, National Space Institute, Technical University of Denmark, Elektrovej 327, 2800 Kgs. Lyngby, Denmark \\
$^{16}$Department of Physics and Astronomy, University of Turku, Vesilinnantie 5, Turku FI-20014, Finland \\
$^{17}$School of Sciences, European University Cyprus, Diogenes Street, Engomi 1516, Nicosia, Cyprus \\
$^{18}$School of Physics \& Astronomy, University of Leicester, University Road, Leicester LE1 7RH, UK \\
$^{19}$School of Physics, Trinity College Dublin, The University of Dublin, Dublin 2, Ireland \\
$^{20}$Instituto de Ciencias Exactas y Naturales (ICEN), Universidad Arturo Prat, Chile \\
$^{21}$National Astronomical Research Institute of Thailand, 260 Moo 4, T. Donkaew, A. Maerim, Chiangmai, 50180 Thailand \\
$^{22}$Armagh Observatory \& Planetarium, College Hill, Armagh BT61 9DG, UK \\
$^{23}$Astrophysics sub-Department, Department of Physics, University of Oxford, Keble Road, Oxford OX1 3RH, UK \\
$^{24}$Astrophysics Research Centre, School of Mathematics and Physics, Queens University Belfast, Belfast BT7 1NN, UK \\
$^{25}$Jodrell Bank Centre for Astrophysics, Department of Physics and Astronomy, The University of Manchester, Manchester M13 9PL, UK \\
$^{26}$School of Physics \& Astronomy, Monash University, Clayton, VIC 3800, Australia \\
$^{27}$Institute of Cosmology and Gravitation, University of Portsmouth, Portsmouth PO1 3FX, UK \\
$^{28}$School of Physics and Astronomy, University of Birmingham, Birmingham, B15 2TT, UK \\
$^{29}$The Oskar Klein Centre, Department of Astronomy, Stockholm University, AlbaNova 106 91, Stockholm, Sweden \\
$^{30}$Department of Particle Physics and Astrophysics, Weizmann Institute of Science, 234 Herzl Street, 7610001 Rehovot, Israel

\bsp	
\label{lastpage}
\end{document}